\documentclass[lettersize,journal,twoside]{IEEEtran}
\usepackage{amsmath,amsfonts}
\usepackage{algorithmic}
\usepackage{algorithm}
\usepackage{array}
\usepackage[caption=false,font=normalsize,labelfont=sf,textfont=sf]{subfig}
\usepackage{textcomp}
\usepackage{stfloats}
\usepackage{url}
\usepackage{verbatim}
\usepackage{graphicx}
\usepackage{cite}
\usepackage{pdfpages}
\usepackage{tabularray}
\UseTblrLibrary{diagbox}
\definecolor{darkblue}{rgb}{0,0,0.5}
\usepackage[colorlinks=true, allcolors=blue, urlcolor=darkblue]{hyperref}
\begin{document}

\title{TsetlinKWS: A 65nm 16.58\textmu W, 0.63mm\textsuperscript{2} State-Driven Convolutional Tsetlin Machine-Based Accelerator For Keyword Spotting}

\author{Baizhou Lin$^{*}$, Yuetong Fang$^{*}$, Renjing Xu, Rishad Shafik,~\IEEEmembership{Senior Member,~IEEE}, Jagmohan Chauhan
        % <-this % stops a space
%\thanks{This paper was produced by the IEEE Publication Technology Group. They are in Piscataway, NJ.}% <-this % stops a space
%\thanks{Manuscript received April 19, 2021; revised August 16, 2021.}}
\thanks{*These authors contributed equally to this work.}
\thanks{Baizhou Lin is with the University of Southampton, SO17 1BJ Southampton, U.K. (e-mail: bzlin713@163.com).}
\thanks{Yuetong Fang, and Renjing Xu are with The Hong Kong University of Science and Technology (Guangzhou), 511453 Guangzhou, China.}
\thanks{Rishad Shafik is with the Microsystems Research Group, School of Engineering, Newcastle University, NE1 7RU Newcastle upon Tyne, U.K.}
\thanks{Jagmohan Chauhan is with the University College London, WC1E 6BT London, U.K.}}
%%%\markboth{Journal of \LaTeX\ Class Files,~Vol.~14, No.~8, August~2021}%
%%%{Shell \MakeLowercase{\textit{et al.}}: A Sample Article Using IEEEtran.cls for IEEE Journals}

%%%\IEEEpubid{0000--0000/00\$00.00~\copyright~2021 IEEE}
\maketitle

\begin{abstract}
The Tsetlin Machine (TM) has recently attracted attention as a low-power alternative to neural networks due to its simple and interpretable inference mechanisms. However, its performance on speech-related tasks remains limited. This paper proposes TsetlinKWS, the first algorithm–hardware co-design framework for the Convolutional Tsetlin Machine (CTM) on the 12-keyword spotting task. Firstly, we introduce a novel Mel-Frequency Spectral Coefficient and Spectral Flux (MFSC-SF) feature extraction scheme together with spectral convolution, achieving a record-high accuracy of 87.35\% for TM-based models on the 12-keyword spotting task. Secondly, we develop an Optimized Grouped Block-Compressed Sparse Row (OG-BCSR) algorithm that delivers a remarkable 9.84$\times$ reduction in model size, significantly improving the storage efficiency of CTMs. Finally, we propose a state-driven architecture tailored for CTMs, which simultaneously exploits data reuse and sparsity to attain high energy efficiency. The full system is evaluated in 65 nm process technology, consuming 16.58 \textmu W at 0.7 V with a compact 0.63 mm\textsuperscript{2} core area. TsetlinKWS demonstrates a Power-Performance-Area (PPA) product of 60.2 and requires only 907k logic operations per inference, corresponding to 1.4$\times$ and 10$\times$ reductions, respectively, compared with state-of-the-art KWS accelerators. These results position the CTM as a highly efficient candidate for ultra-low-power speech applications.

\end{abstract}


\section{Introduction}
\IEEEPARstart{T}{he} Tsetlin Machine (TM) is an emerging machine learning algorithm founded on propositional logic and learning automata. It operates on binary inputs and performs logical pattern matching for inference. Different from neural networks (NNs), its logic-based and multiplication-free computational paradigm offers greater suitability for low-power edge deployment. Furthermore, its straightforward state update strategy without backpropagation makes it well-suited for on-device learning. These characteristics of the TM are essential for edge speech applications, e.g., keyword spotting (KWS), which have stringent requirements for low power consumption, low latency, and on-device adaptability.

\begin{figure}[t]
    \centering
    \includegraphics[width=0.9\linewidth]{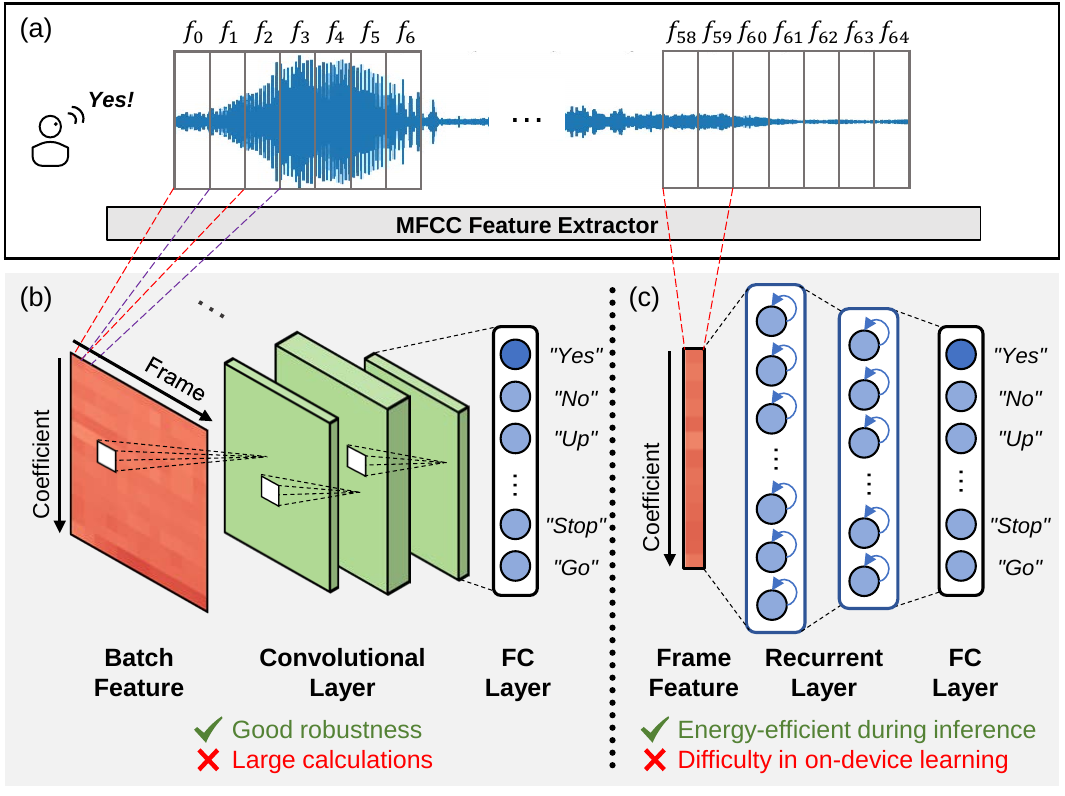}
    \caption{NN-based KWS accelerator categories. (a) Audio signal. (b) Batch-based accelerators are represented by convolutional neural networks \cite{liu202022nm},\cite{liu20233},\cite{lin202316},\cite{bernardo2020ultratrail},\cite{fu202595},\cite{li20262}. They perform inference in a multi-frame manner because neurons do not have states. (c) Frame-based accelerators are represented by recurrent neural networks \cite{frenkel2022reckon},\cite{chong20212},\cite{giraldo2020vocell},\cite{chen2024deltakws}. They process only one frame of features per inference.}
    \label{fig: 1}
\end{figure}

Existing NN-based KWS accelerators follow either the batch-based or frame-based processing paradigm, as shown in Fig.~\ref{fig: 1}. Although batch-based accelerators can offer improved robustness \cite{liu202022nm}, KWS chips supporting on-chip learning still incur substantial power consumption \cite{qian2025chip},\cite{zheng2019ultra},\cite{chiang202214}. Similarly, while frame-based accelerators, typically built around recurrent neural networks (RNNs), enable energy-efficient inference by reusing hidden states across time and reducing redundant computation, training multi-layer RNNs requires intermediate activations and hidden states to be stored across time steps \cite{frenkel2022reckon}, resulting in substantial memory overhead that limits their suitability for on-device learning. The simple inference and learning mechanisms of TMs present a new solution for this challenge. Previous studies have shown that TMs achieve competitive accuracy and energy benefits compared with NNs on simple datasets \cite{rahman2024matador}, and also exhibiting a clear advantage in on-device learning \cite{tunheim2024tsetlin},\cite{mao2025dynamic}. However, two current challenges hinder their widespread adoption in edge deployment: firstly, for more complex tasks such as speech recognition, their performance remains limited relative to NN-based methods; secondly, there is still a lack of a hardware architecture tailored for the Convolutional Tsetlin Machine (CTM) to fully leverage its sparsity advantage while maintaining efficient data reuse. \IEEEpubidadjcol 

In this work, we focus on the KWS problem to explore the limitations of TMs on speech tasks. KWS is a representative low-power edge application, widely used to enable wake-up and command functions. Its always-on mode and tight energy budgets impose particularly stringent requirements on latency, power consumption, and model efficiency, making it a critical test case for TM-based methods. However, existing approaches that apply the Vanilla TM to KWS have failed to deliver competitive accuracy. This limitation arises because Vanilla TMs rely on long-term pattern matching over Boolean features, which hampers the model’s generalization ability. This tendency is evident in experimental results: Lei et al.~\cite{lei2021low} observed that increasing the number of target keywords from six to nine led to a considerable accuracy reduction, from 86.5\% to 80.5\%. Subsequent work reveals a similar issue: although REDRESS introduced a compression-aware training method and attained 87.1\% accuracy, this gain was confined to a simplified 6‑keyword task~\cite{maheshwari2023redress}. There remains a persistent difficulty in scaling TMs to more complex scenarios.

Beyond the algorithmic limitations of TM in speech tasks, a second critical challenge lies in the architectural mismatch between existing accelerators and the computational paradigm of the CTM. Current CTM-based accelerator designs are largely inherited from Convolutional Neural Network (CNN) accelerators, which are unable to efficiently achieve both data reuse and sparse utilization under the constraints of NN computing paradigms. As a result, the exceptionally high bit-level sparsity and deterministic sparsity patterns of CTMs cannot be fully utilized. Tunheim et al.~\cite{tunheim2025all} proposed a CTM image classification accelerator that does not support sparse utilization. The ineffective computation brought by its CNN-style dense architecture leads to irrational bandwidth utilization; that is, due to high concurrency requirements, the model weights can only be stored in registers. To overcome these inefficiencies, we systematically analyze the computational patterns of CTMs and the design principles of existing edge Artificial Intelligence (AI) accelerators, leading us to propose a new architecture that simultaneously enables sparse acceleration and efficient data reuse.

In this paper, we address the challenges of low accuracy and limited hardware efficiency of CTMs on speech tasks by designing a KWS inference-only accelerator. Firstly, for the feature processing, we introduce a Mel-Frequency Spectral Coefficient and Spectral Flux (MFSC-SF) algorithm to enhance feature quality, and a streaming threshold update circuit for efficient feature binarization. Secondly, to address storage constraints, we apply block and grouped compressions, formulated as a maximum weight matching problem, for finding the optimal compressed model size. Thirdly, we propose a state-driven architecture that enables sparse acceleration and efficient data reuse. Finally, we adopt a two-stage simulated annealing algorithm for balanced clause scheduling. The main contributions of this work are summarized as follows:

\begin{itemize}
    \item \textbf{Algorithmic Improvement for CTM-based KWS:} We introduce a novel MFSC-SF algorithm and spectral convolution operation for CTMs, achieving 87.35\% accuracy on the 12-keyword spotting task. This work significantly narrows the accuracy gap between TMs and edge-deployed NNs, demonstrating the feasibility of logic-based models for complex speech recognition.
    \item \textbf{CTM-Specific Accelerator:} We design an efficient state-driven accelerator tailored for CTMs that unlocks high data reuse and full sparsity utilization. The proposed accelerator achieves a remarkably low power consumption of 16.58 \textmu W within a compact core area of 0.63 mm$^2$, and delivers a competitive Power-Performance-Area (PPA) product of 60.2 among NN-based KWS accelerators.
    \item \textbf{Edge-Optimized Deployment:} We propose an Optimized Grouped Block-Compressed Sparse Row (OG-BCSR) algorithm and a two-stage scheduling scheme, achieving a 9.84$\times$ model size reduction and a 12.2\% PE utilization improvement.

\end{itemize}

The rest of this paper is structured as follows: Section~\ref{sec: 2} reviews the background of CTM and its associated hardware\slash algorithm design challenges; Section~\ref{sec: 3} details the proposed state-driven CTM-based KWS system; Section~\ref{sec: 4} presents and analyzes the implementation results; Finally, we summarize our work in Section~\ref{sec: 5}.

\section{Background and Challenges}\label{sec: 2}
\subsection{Compression for Tsetlin Machines}\label{subsec: 2.1}
\begin{figure}[t]
    \centering
    \includegraphics[width=0.9\linewidth]{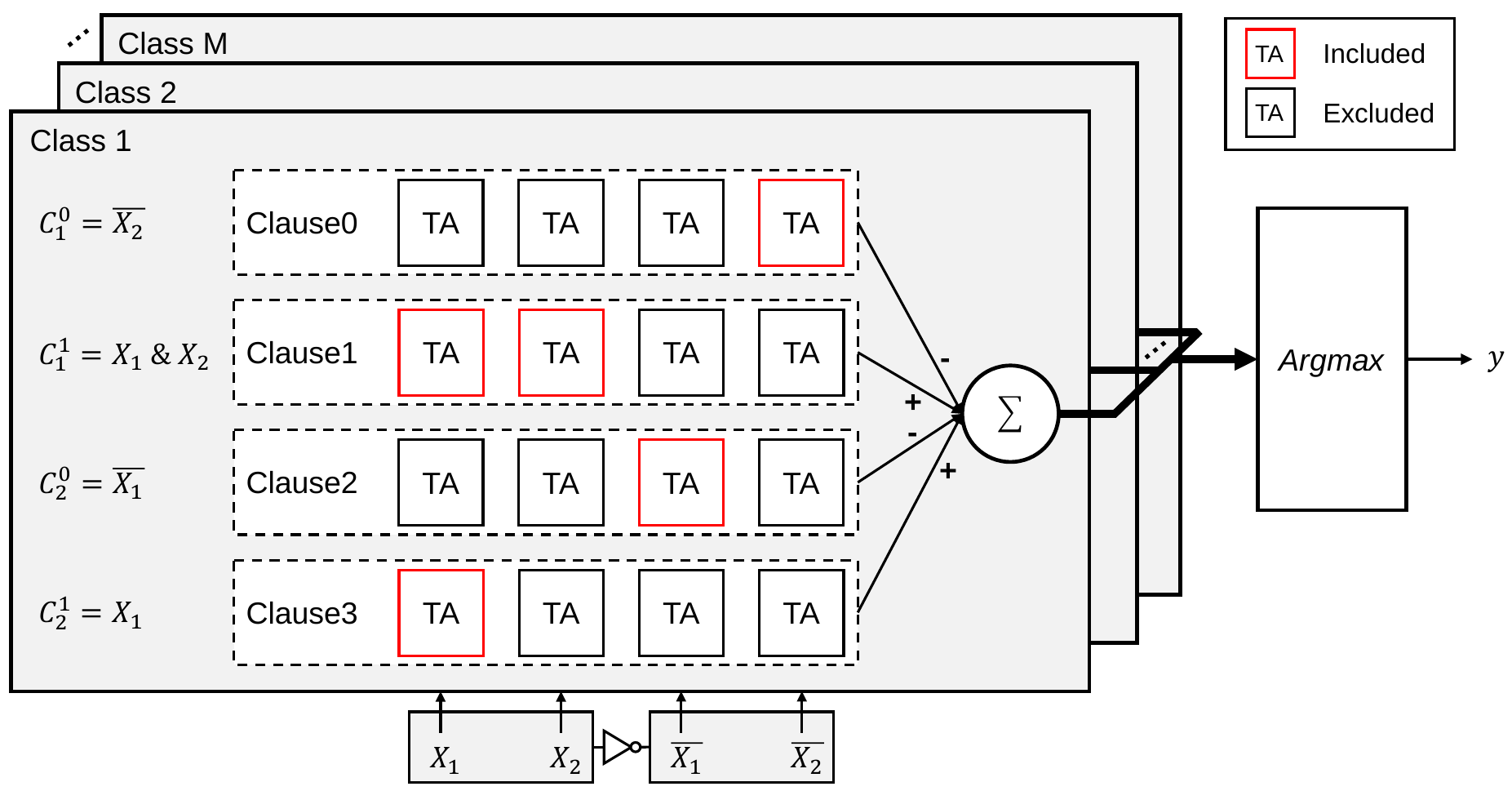}
    \caption{The structure of the Tsetlin Machine. $X_k$ refers to the Boolean input. Included TAs and excluded TAs determine the matching pattern of clause $C_i^j$.}
    \label{fig: 2}
\end{figure}

The TM is a logic-based learning algorithm that uses a set of logical clauses to capture discriminative patterns in data. As shown in Fig.~\ref{fig: 2}, the input to a TM is a set of Boolean literals, namely variables and their negations. Each clause consists of multiple Tsetlin Automata (TAs) operating independently to decide whether a literal should be included or excluded from the clause. The states of TAs (included/excluded) are optimized during training according to the hyperparameters $T$ and $s$, with further details on these hyperparameters provided in~\cite{granmo2018tsetlin}. Through this collective decision-making, the clause converges to a Boolean expression that captures a specific discriminative pattern in the data. When a Boolean expression is satisfied, the weight associated with the corresponding clause is added to the confidence score of the target class~\cite{abeyrathna2021extending}. Ultimately, after processing all clauses of each class, the argmax unit identifies the class with the highest confidence as the prediction result. Similar to CNNs, CTMs~\cite{granmo2019convolutional} introduce a convolutional mechanism that enables parameter sharing by applying the same clauses across multiple input regions. Specifically, as long as the convolution kernel is aligned with the target pattern in any sliding window, the corresponding clause is satisfied. For trained TMs, a key characteristic is their extreme sparsity, with included TAs accounting for only about 1\%, resulting in a highly condensed representation.

This high bit-sparsity presents a critical opportunity for compressing TM models, which is particularly desirable for edge deployment. However, existing methods often apply compression algorithms directly to the model parameters, without introducing special treatments for the extreme sparsity. For example, Bakar et al.~\cite{bakar2022logic} utilized run-length encoding (RLE) to achieve a 98.68\% compression rate on a 6-keyword spotting task, while Maheshwari et al.~\cite{maheshwari2023redress} proposed include-encoding with compression-aware training to reduce model size from 1325.4 KB to 15.4 KB on the same task. We observe that an extremely high sparsity rate, e.g., 99\%, can have a counterproductive effect on compression efficiency. To address this issue, we propose block compression and grouped compression schemes to achieve efficient compression, which substantially reduce the average storage overhead per class compared to prior approaches.

\subsection{Architectural Dilemma}\label{subsec: 2.2}
\begin{figure}[t]
    \centering
    \includegraphics[width=0.9\linewidth]{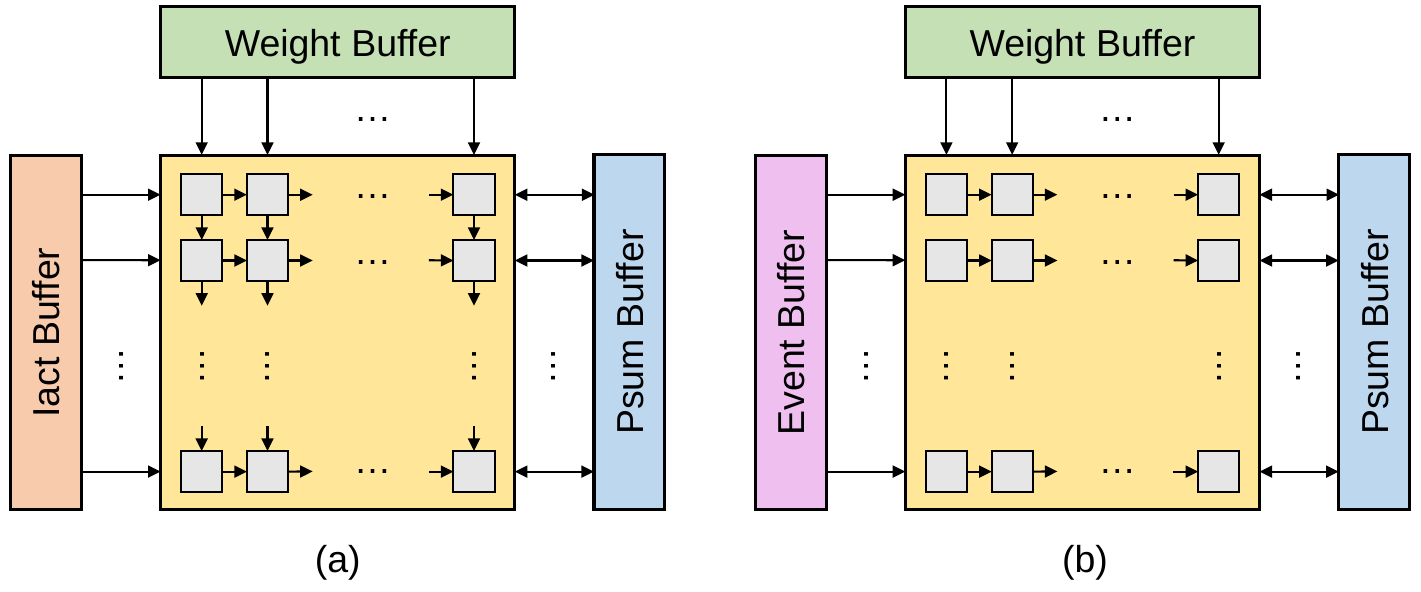}
    \caption{Two common energy-efficient edge accelerator architectures. (a) Data-driven architecture: efficient for data reuse but inflexible for sparse acceleration. (b) Event-driven architecture: efficient for sparse acceleration but limited in data reuse.}
    \label{fig: 3}
\end{figure}

Previous CTM accelerators often directly followed NN-based accelerators without a systematic analysis of their underlying design principles, leading to computational inefficiency. Indeed, the properties of the CTM enable us to overcome the limitations of NN-oriented accelerator designs and achieve greater efficiency. In edge AI accelerators, data reuse and sparsity utilization are critical for improving energy efficiency, while simultaneously achieving both goals has remained challenging under NN computing paradigms. As illustrated in Fig.~\ref{fig: 3}, two common architectures exemplify this trade-off.

Data-driven architectures, commonly employing systolic arrays, leverage the regular dataflow of Artificial Neural Networks (ANNs) to maximize data reuse and parallelism \cite{yin20171}. Specifically, weights and input activations (iacts) are propagated simultaneously to realize data reuse. However, sparse ANNs \cite{han2015learning} cannot directly benefit from this hardware due to irregular dataflow. Although several works \cite{zhang2016cambricon},\cite{parashar2017scnn},\cite{deng2021gospa} have proposed solutions for efficient sparse utilization, this inherent limitation results in additional hardware overhead, such as expensive index modules and intersection strategies. Therefore, sparse utilization is difficult to achieve efficiently at a low cost in data-driven architectures. 

From the perspective of high sparsity, another model more similar to TMs is the Spiking Neural Network (SNN), which has spurred the development of event-driven architecture. This architecture represents a new design paradigm that achieves sparsity acceleration by processing only the data required for spiking events. However, because spiking events are unpredictable and unstructured, weights are difficult to reuse vertically. For instance, SpinalFlow \cite{narayanan2020spinalflow} supports only unrolling output channels in a 1$\times$128 Process Element (PE) array due to the inability to reuse weights between spiking events. Some works \cite{sommer2022efficient}, \cite{plagwitz2023spike} have constructed event-driven accelerators and compared them with FINN \cite{blott2018finn}. The results also show that inefficient data reuse greatly affects the energy efficiency of event-driven accelerators.

We found that the limited benefit of data-driven accelerators supporting sparse acceleration stems from the modest sparsity rate and multi-bit data, while the underwhelming benefit of event-driven accelerators arises from uncertain spiking events. Fortunately, CTMs are highly sparse, logic-based, and sparse-deterministic, which motivates us to propose a state-driven architecture that simultaneously achieves sparse utilization and efficient data reuse, making it particularly suitable for low-power edge applications.

\subsection{Performance and Benchmarking in KWS}
The keyword spotting accelerator typically consists of a front-end feature extractor and a back-end classifier. 

To reduce the power consumption of the KWS system, the feature extractor must generate compact yet discriminative representations to reduce the computational burden of the back-end classifier while maintaining accuracy. Related research has shown that binary features are sufficient for KWS tasks \cite{lin202316},\cite{lei2021low}. Jiao et al. \cite{jiao2022thermometer} utilized a Log Mel-Frequency Spectral Coefficient (MFSC) algorithm with fixed thermometer codes to introduce Boolean inputs in the early stage for Binary Neural Networks (BNNs), alleviating computational demands significantly. However, existing binarization methods often rely on fixed thresholds, making them vulnerable to sensor degradation and environmental variations. Additionally, directly using MFSC as the input feature for logic-based models (i.e., TMs) will lead to accuracy loss. To address these issues, we propose an improved feature extraction algorithm with a dynamic binarization method to enhance performance and robustness.

The back-end classifier can be categorized into two types in terms of their computational paradigms, as illustrated in Fig.~\ref{fig: 1}. Batch-based accelerators process a long-term window of features, incurring a higher computational cost (e.g., 3 million operations per inference \cite{liu202022nm}) while exploiting longer temporal contexts for improved noise robustness. In contrast, frame-based accelerators operate in a streaming fashion, leveraging the stateful nature of RNNs to process features frame by frame. This paradigm avoids repeatedly processing long feature windows by maintaining temporal states, making it an attractive option for energy-constrained speech applications. For instance, Chen et al. \cite{chen2024deltakws} proposed a delta-encoder module to differentiate input features and generate sparse events to drive RNN model computation, requiring only 15k operations per inference. To provide a more comprehensive comparison across these two accelerator paradigms, we introduce an inference-core PPA product metric that jointly captures power consumption, performance, and silicon area. Because KWS inference is typically latency-sensitive, inference latency is adopted as the performance metric. The normalized inference-core PPA product is defined as

\begin{equation}
\label{eq:ppa}
%\mathrm{PPA~Product} = P \times L \times A.
\mathrm{PPA~Product} = \underbrace{P \times L}_{\mathrm{Energy}} \times A.
\end{equation}
where $P$ denotes the normalized inference-core power, $L$ denotes the inference latency, and $A$ denotes the normalized inference-core area. Since energy is determined by power and execution time, $P \times L$ represents the normalized inference-core energy. The PPA product serves as a holistic metric for accelerator evaluation because techniques for optimizing energy consumption often incur non-trivial silicon area overheads. Consequently, a lower PPA product indicates a superior balance among these multi-dimensional design objectives.

While TMs are often claimed to be efficient \cite{lei2020arithmetic}, comparisons are frequently made against non-optimized NNs. A more rigorous evaluation, however, requires benchmarking state‑of‑the‑art (SoTA), highly optimized silicon implementations. Our TsetlinKWS accelerator belongs to the batch-based accelerator category. In this work, we not only demonstrate the performance of CTMs in speech applications but also aim to showcase their competitive PPA products with frame-based accelerators when crossing classification boundaries.

\section{State-Driven Convolutional Tsetlin Machine KWS Accelerator}\label{sec: 3}

In this section, we elaborate on the end-to-end design of the proposed system. Section \ref{subsec: 3.1} outlines the overall architecture of TsetlinKWS. Section \ref{subsec: 3.2} details the algorithm and hardware implementation of the MFSC-SF feature extractor. The OG-BCSR compression algorithm is introduced in Section \ref{subsec: 3.3}. Finally, Section \ref{subsec: 3.4} explains the state-driven strategy.

\subsection{Architecture Overview}\label{subsec: 3.1}
\begin{figure}[t]
    \centering
    \includegraphics[width=1.0\linewidth]{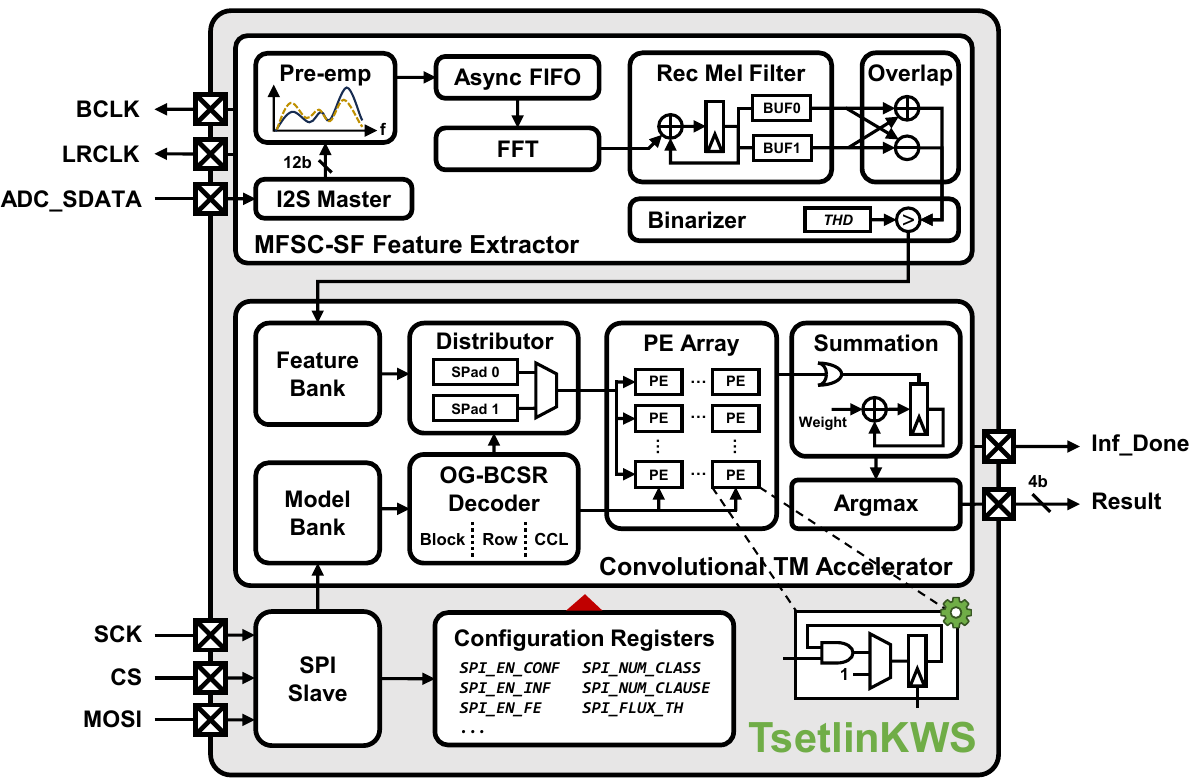}
    \caption{The system architecture of TsetlinKWS.}
    \label{fig: 4}
\end{figure}

Fig.~\ref{fig: 4} illustrates the system architecture of the proposed TsetlinKWS. TsetlinKWS primarily consists of three components: the Serial Peripheral Interface (SPI) configuration interface, an MFSC-SF feature extractor, and a convolutional TM accelerator. The SPI interface is used for chip configuration and model loading. Audio data is received through the Integrated Interchip Sound (I2S) interface. The feature extractor is used for extracting and binarizing input speech features. The CTM accelerator is constructed to achieve efficient sparse dataflow utilization by algorithm-hardware co-design. Specifically, the OG-BCSR decoder enables horizontal data reuse of feature data through synchronous decompression. The distributor transmits data to the subsequent PE array according to the decompression signal. The PE array is organized as a 58$\times$5 spatial architecture to balance throughput and hardware complexity. In this configuration, the dimension of 58 naturally aligns with 58 convolutional sliding windows, while the choice of 5 columns provides sufficient acceleration yet avoids the memory partitioning overhead associated with higher parallelism. When a batch of clauses is completed, clause results are serially transferred to the summation module for confidence updating. Once all clauses for all classes are completed, the argmax module updates the inference result and synchronously sends an interrupt signal to the external host. TsetlinKWS is an open-source accelerator available at: \url{https://github.com/Baizhou-713/TsetlinKWS}.

\subsection{MFSC-SF Feature Extractor}\label{subsec: 3.2}
\begin{figure}[t]
    \centering
    \includegraphics[width=1\linewidth]{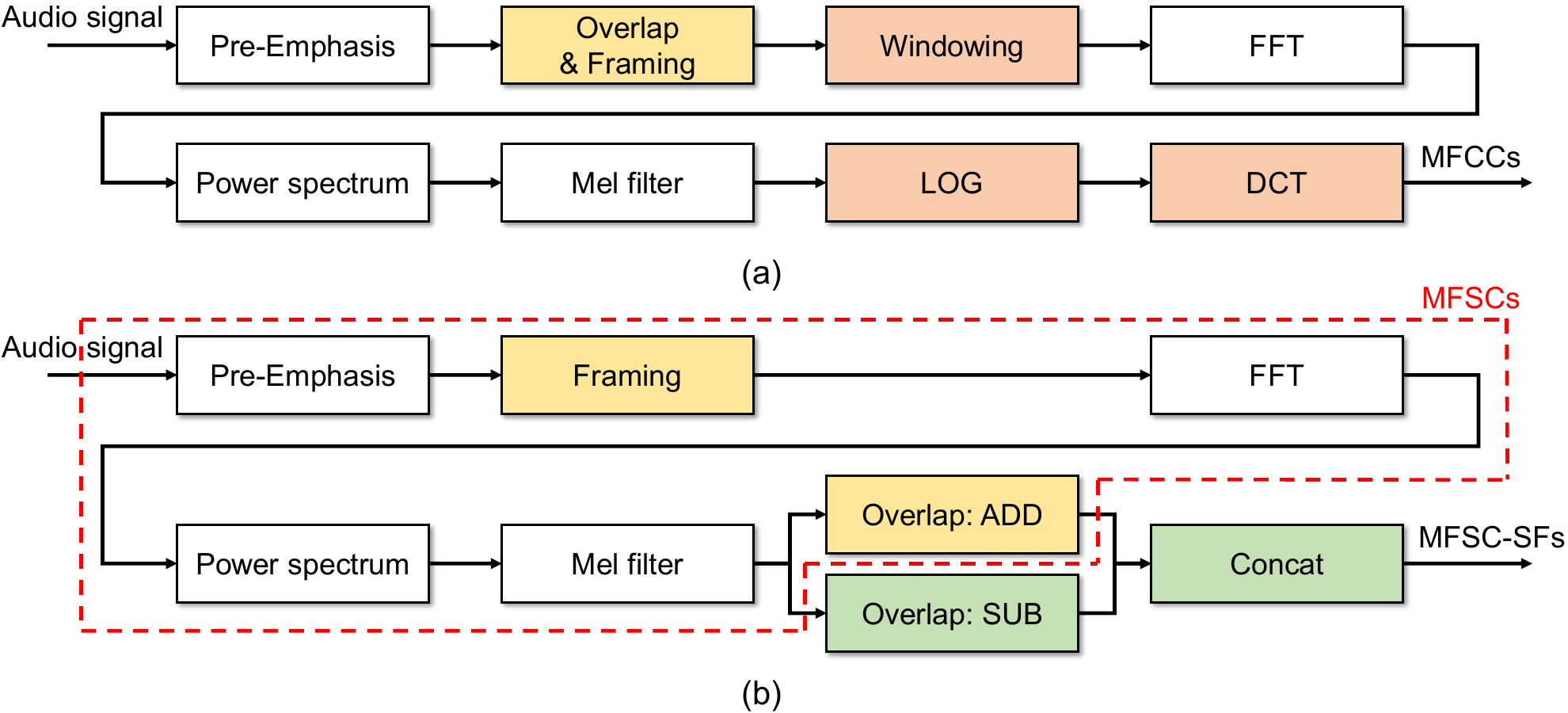}
    \caption{Feature extraction process: (a) Traditional MFCC flow. (b) The proposed MFSC-SF flow. An additional overlap difference is proposed to achieve the calculation of spectral flux, and some sub-modules are deleted.}
    \label{fig: 5}
\end{figure}

\subsubsection{Overview}
As illustrated in Fig.~\ref{fig: 5}(a), the conventional Mel-Frequency Cepstral Coefficients (MFCC) extraction pipeline begins with a pre‑emphasis filter to compensate for spectral tilt and amplify high‑frequency components. The signal is then segmented into overlapped frames and multiplied by a window function, typically with length $N$ and 50\% overlap. An $N$-point FFT converts each frame to the frequency domain, from which the power spectrum is computed, yielding $N/2+1$ coefficients. These are passed through a Mel-scale filterbank to produce $M$ Mel-spectral coefficients, where $M$ typically ranges from 26 to 32. Logarithmic (LOG) compression followed by a discrete cosine transform (DCT) is then applied to obtain the MFCC representation.

The proposed MFSC‑SF algorithm builds upon the work in \cite{han2006efficient}. As illustrated in Fig.~\ref{fig: 5}(b), the overlap operation is deferred until after the Mel filterbank, where the effect is achieved by summing the MFSCs of adjacent frames. This modification reduces the subframe length by half, thereby lowering buffer and FFT resource requirements. In addition, several power‑saving strategies are adopted: (1) elimination of the windowing operation, (2) approximation of the power spectrum by absolute values, and (3) substitution of triangular Mel filters with rectangular filters. These simplifications incur less than 1\% accuracy loss while significantly reducing energy consumption.

To make spectral features easier to extract by CTMs, several additional changes are introduced in the MFSC-SF flow. We remove the LOG and DCT operations for two reasons. First, the single‑layer CTM model demonstrates limited ability to learn discriminative features after nonlinear transformations; second, the DCT will convert the feature to a non‑local feature, which conflicts with the proposed spectral convolution (Section \ref{subsec: 3.4}). Notably, these operations are also commonly bypassed in neural network KWS accelerators for power efficiency. Furthermore, distinct from the conventional MFCC flow, we introduce Spectral Flux (SF) to improve accuracy for CTMs. While SF is traditionally defined as the FFT difference between adjacent frames, here it is repositioned after the Mel filter to align with the MFSC, as shown in \eqref{sf}. The final feature representation is thus obtained by concatenating the MFSC and SF components.

\begin{equation}
\label{sf}
SF(t) = |Mel(t) - Mel(t - 1)|.
\end{equation}

\begin{figure*}[t]
    \centering
    \includegraphics[width=0.95\linewidth]{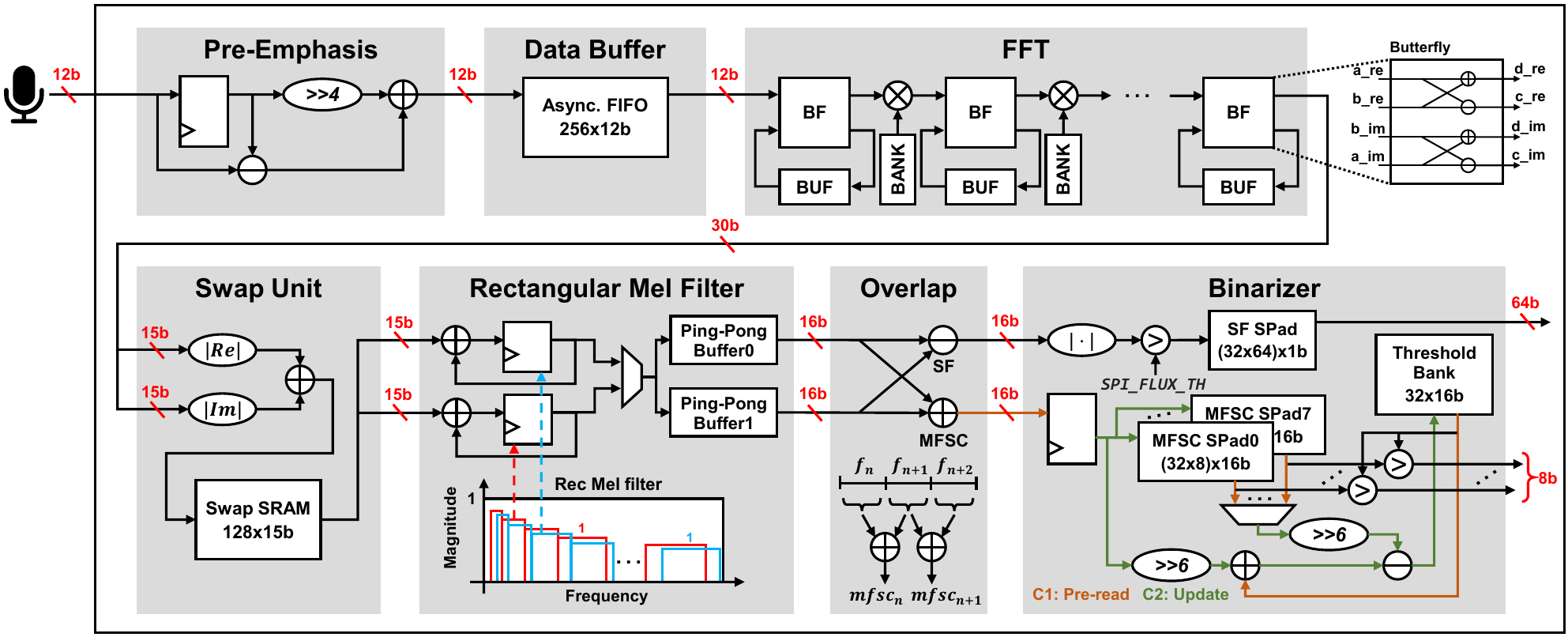}
    \caption{The architecture diagram of the MFSC-SF feature extractor. MFSC and SF are obtained in the overlap unit and then are binarized in the binarizer. MFSC is binarized using the dynamic mean, while SF is binarized using a fixed threshold.}
    \label{fig: 6}
\end{figure*}

The theoretical motivation for explicitly introducing SF is driven by the computing paradigm disparity between TMs and ANNs. Unlike neural networks, which can easily learn to extract temporal feature differences via synaptic weights (e.g., $w_1 = -1, w_2 = 1$), TMs fundamentally rely on propositional logic and cannot efficiently perform arithmetic deductions. Therefore, offloading this arithmetic burden to a simple subtraction operation to provide a critical prior is essential. By projecting both MFSC and SF into the Boolean space as complementary features, CTMs can efficiently recognize both stable acoustic patterns and transient acoustic events, thereby overcoming their limitations and improving performance.

\subsubsection{Architectural Design}
Fig.~\ref{fig: 6} illustrates the overall architecture of the proposed feature extractor.  The pre‑emphasis module follows the design in~\cite{liu202022nm}. Each subframe has a fixed length of 256 samples and is buffered using an asynchronous first‑in‑first‑out (FIFO) buffer, with the frame clock serving as the write clock and the system clock as the read clock.  A 256‑point FFT is implemented using an eight‑stage radix‑2 single‑path delay feedback (R2‑SDF) architecture. To reduce power and storage overhead, stage‑wise quantization is applied: the integer bit widths are [12, 13, 13, 13, 14, 14, 14, 15], while the fractional bit widths are [1, 1, 1, 1, 1, 0, 0, 0], resulting in a 15‑bit FFT output. For a 256‑point real FFT, 129 spectral coefficients are retained; one coefficient is discarded for hardware simplification without accuracy degradation.
The Mel filter unit employs two registers to compute odd and even filter outputs separately. By adopting rectangular filters with unity gain, multiplication operations are entirely eliminated. A ping‑pong buffer stores the results of adjacent frames, enabling overlap processing. SF is then computed in the overlap unit using only a subtractor, yet it provides substantial accuracy improvements. Finally, a binarizer converts the 16‑bit outputs into compact binary feature representations.

In this work, we design a binarization circuit for CTM-based KWS. Since spectral flux reflects the variation between adjacent frames, a fixed threshold method is adopted for SF binarization. In this implementation, the SF threshold is empirically set to 1024 through the SPI configuration interface. The SF scratch pad (SPad) stores the Boolean values of 64 frames and forwards them to the downstream feature bank. Unlike SF, MFSC represents the energy level of the current frame, which is sensitive to microphone gain variations. As a result, a fixed threshold cannot satisfy practical deployment requirements. Previous TM works typically employ quantile binning for binarization, yet it introduces considerable quantization errors because audio features usually exhibit highly imbalanced distributions. Therefore, a dynamic average thresholding circuit is proposed to realize streaming threshold updates. To eliminate the costly division operation in average calculation, the batch size is set to 64, so that shifting can be used to approximate the division operation.

The threshold update process is executed in two cycles. In the first cycle, the original threshold is read from the threshold bank, and the new MFSC value is stored in the register. Simultaneously, the old MFSC value is read from the MFSC SPad according to the location where the new MFSC value is to be written. In the next cycle, the new threshold is obtained by adding the new MFSC value to the original threshold and subtracting the old MFSC value. For each inference, all data stored in the MFSC SPad is read out for binarization. To improve parallelism, the MFSC SPad is divided into eight banks, allowing eight values to be read simultaneously. This binarization method allows all memories to be mapped to single-port Static Random-Access Memory (SRAM) macros.

\begin{figure}[t]
    \centering
    \includegraphics[width=1\linewidth]{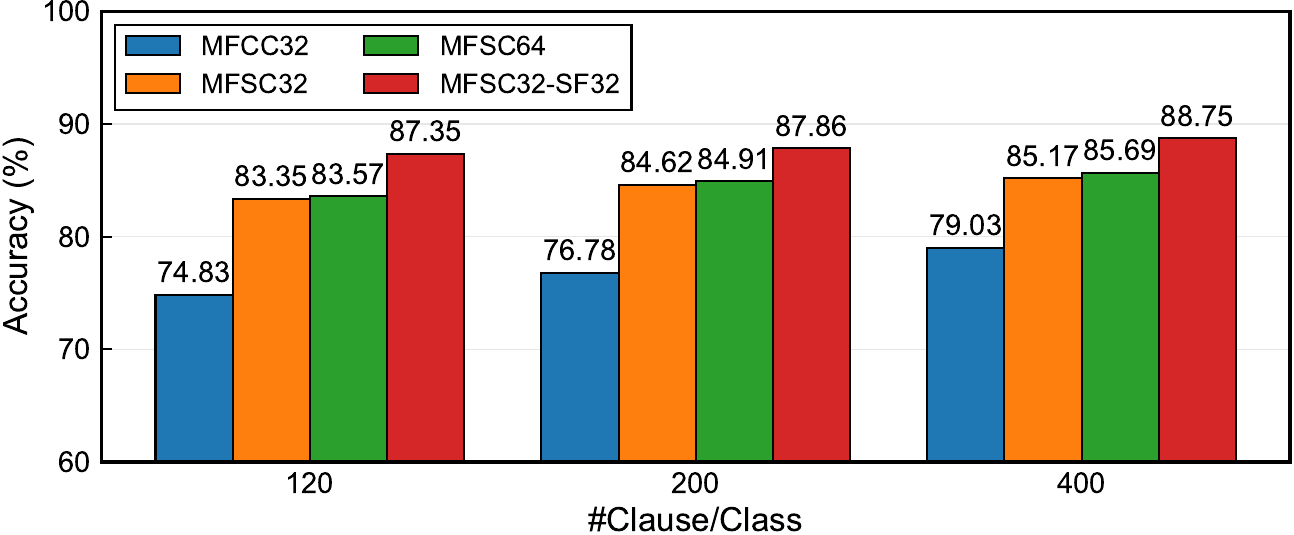}
    \caption{Accuracy of CTMs under different pre-processing algorithms and at different model scales. ``32/64" represents the number of filter bank channels. The 120-Clause CTM with MFSC32-SF32 is the selected model in this work.}
    \label{fig: 7}
\end{figure}

We evaluate MFCC, MFSC, and MFSC-SF with varying coefficient sizes for feature extraction and train CTM models at different scales, as shown in Fig.~\ref{fig: 7}. Compared with MFSC using 32 coefficients (MFSC32), MFSC-SF improves the accuracy of the 120-clause CTM by 4 percentage points. Furthermore, consistent with the theoretical analysis presented earlier, MFSC achieves higher accuracy than MFCC by eliminating the LOG and DCT operations. Due to the computational paradigm of the convolutional model, the input feature dimension does not significantly affect the model size. However, the benefits of introducing the MFSC-SF algorithm allow the model to be scaled down to a smaller size without compromising accuracy.

\subsection{OG-BCSR Algorithm}\label{subsec: 3.3}
\subsubsection{Overview}

In this work, we choose a 64$\times$7 convolution kernel. The details about the selection of the convolution kernel size are presented in the Appendix. Since the convolution kernel of CTM contains positional TAs, we concatenate them into the matrix. To account for both positive and negative literals, the resulting TA action matrix has dimensions of 64$\times$16. For the 120-Clause CTM model, the size of the TA action matrix is 180 KB. To achieve efficient compression, we introduce an Optimized Grouped Block-Compressed Sparse Row algorithm, which maximizes the compression rate under given constraints. This algorithm is proposed based on the following observation: in highly sparse matrices, empty rows can diminish the compression efficiency of the standard Compressed Sparse Row (CSR) algorithm. We improve the compression rate through two methods: block compression and grouped compression. The purpose of block compression is to skip the unnecessary list storage using a block index list. The block index indicates whether a block contains any included TAs, with every two rows combined into a single block. The purpose of grouped compression is to reduce the sparsity rate of the matrix. Therefore, an additional clause index list is added to the list to indicate which clause the included TA comes from.
  
\begin{figure}[t]
    \centering
    \includegraphics[width=0.8\linewidth]{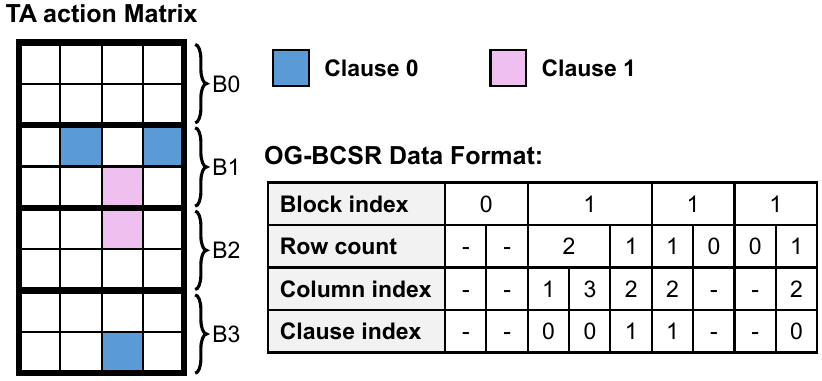}
    \caption{Example of compressing the TA action matrix using OG-BCSR. The block index is used to skip storage of empty blocks, e.g., the first block B0 is empty. The row count indicates the number of included TAs in each row. The column index indicates their column addresses.}
    \label{fig: 8}
\end{figure}

Fig.~\ref{fig: 8} shows an example of OG-BCSR compression. If the block is empty, storage of the subsequent list is skipped. Unlike the original CSR algorithm~\cite{filippone2017sparse}, we use direct row index storage instead of pointer storage. This is because the included TA only needs to be accessed sequentially and does not need to perform multiplications with other matrices. The use of direct storage can reduce the overhead of the decompression circuit.

\subsubsection{Graph-Based Grouping}

In the OG-BCSR algorithm, the grouping strategy directly influences the compression outcome. We observe that this problem can be transformed into a maximum weight matching problem in graph theory to obtain the optimal grouping scheme. This is because in the OG-BCSR compression list, the length of the block index list is fixed, while the lengths of the column index list and clause index list are determined by the number of included TAs. These lists cannot be optimized. Therefore, the optimal compression is achieved when the length of the row count list is minimized. Fortunately, the length of the row count list is positively correlated with the number of non-empty blocks, which can be known in advance before compression.

\begin{figure}[t]
    \centering
    \includegraphics[width=0.8\linewidth]{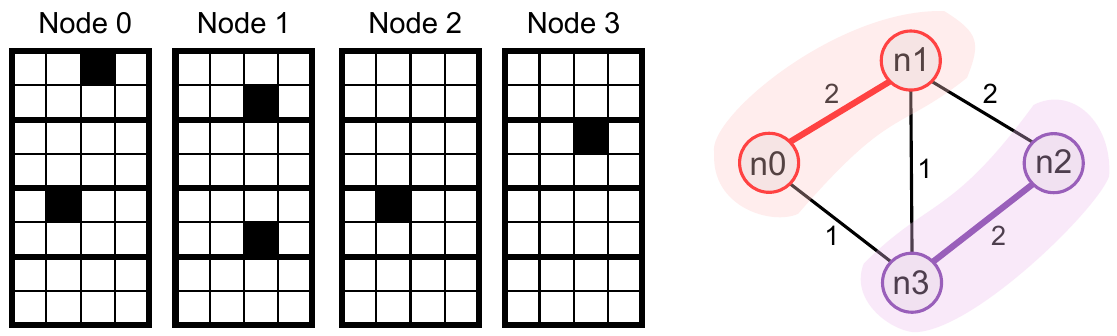}
    \caption{Construction process from the TA action matrix to the graph. The weight of edges is the number of empty blocks after merging. In the example, n0-n1 and n2-n3 form the optimal matching.}
    \label{fig: 9}
\end{figure}

An example of converting matrix grouping into a graph is depicted in Fig.~\ref{fig: 9}, where each TA action matrix is treated as a node (left), and nodes are connected pairwise to form a general graph (right). We define two constraints: if merging two matrices would result in overlapping non-zero elements or exceed the maximum allowable row count, they are considered to have no connecting edge, such as node 0 and node 2. The optimization goal is to find a matching between nodes such that the total weight of the connecting edges between each pair of nodes is maximized, i.e., the scheme with the maximum number of empty blocks after merging is obtained. This maximum weight matching problem is solved using the Blossom algorithm \cite{galil1986efficient}, which efficiently identifies augmenting paths in general graphs.

\begin{figure}[t]
    \centering
    \includegraphics[width=0.8\linewidth]{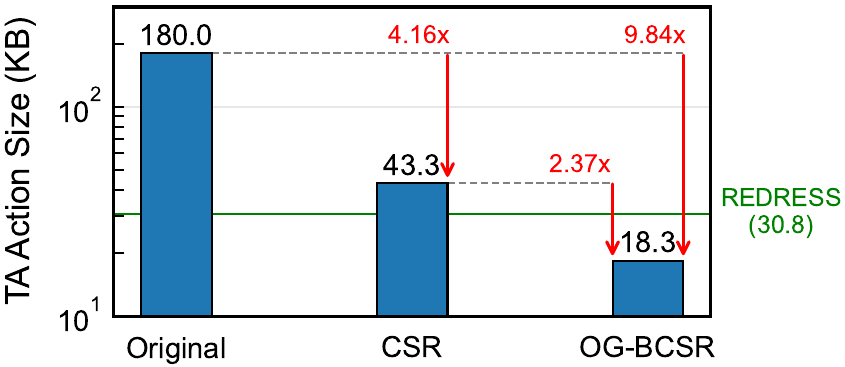}
    \caption{Compression results of CSR and OG-BCSR algorithms.}
    \label{fig: 10}
\end{figure}

We restrict the number of included TAs per row after merging to no more than 7. Due to the low similarity and high sparsity of the matrix, this constraint is violated in only a few cases, which does not affect the optimal solution. Therefore, the row count list can be represented by 3 bits to reduce storage overhead. In this work, the total number of included TAs in the 120-Clause CTM model is 15.6k. After constructing the general graph, the maximum weight matching algorithm is applied to obtain the optimal grouping. The result is shown in Fig.~\ref{fig: 10}, where the proposed OG-BCSR algorithm can further reduce the model size by 2.37$\times$ compared to CSR. Since the SoTA compression work, REDRESS \cite{maheshwari2023redress}, focuses on a 6-keyword task, we linearly scale its memory requirements to align with our task. The result demonstrates that our algorithm achieves a lower memory footprint.

\subsection{Algorithm-Hardware Co-design}\label{subsec: 3.4}

In this work, the algorithm-hardware co-design primarily consists of three parts: the spectral convolution operation, a state-driven architecture, and a two-stage scheduling optimization algorithm, followed by a theoretical performance analysis.

%\subsubsection{Spectral Convolution Kernel}
\subsubsection{Spectral Convolution Operation}

\begin{figure}[t]
    \centering
    \includegraphics[width=0.9\linewidth]{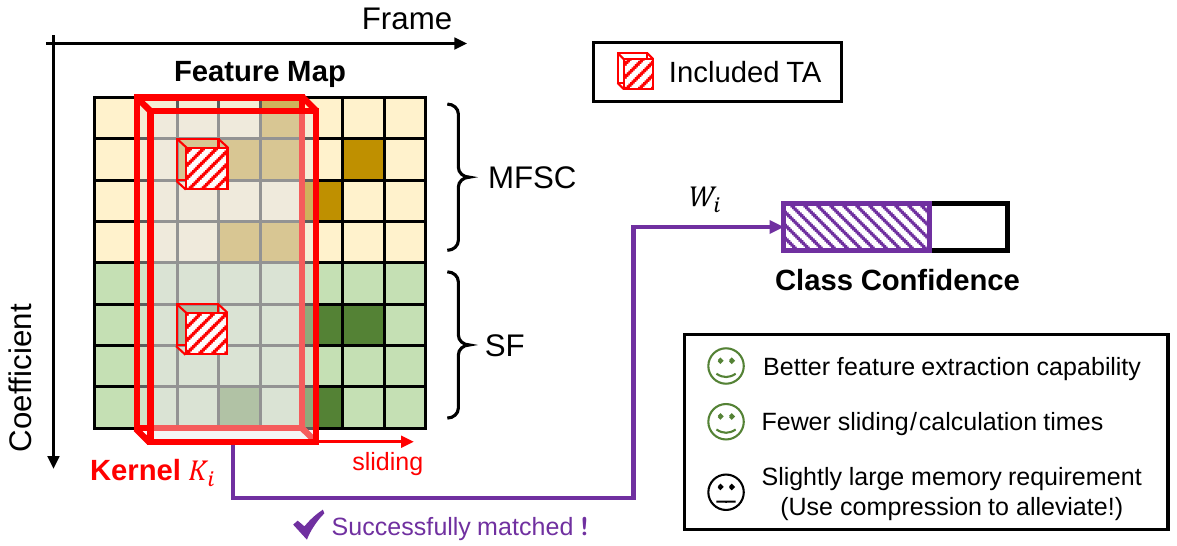}
    \caption{The spectral convolution kernel covers the entire spectral coefficient dimension to achieve efficient feature matching. Two smiling faces denote the advantages of the spectral convolution kernel in terms of performance and power, while a neutral face represents an acceptable area trade-off. $W_i$ refers to the weight of the convolution kernel $K_i$.}
    \label{fig: 11}
\end{figure}

In CNNs, small convolution kernels are commonly used in multi-layer networks to accomplish specific recognition tasks. However, the current CTM remains a single-layer model, and related research \cite{lei2021low} indicates that one Boolean value per coefficient is sufficient to represent most information. The use of small kernels will prevent the CTM model from having a sufficiently large receptive field to recognize global spectral features. Therefore, the spectral convolution kernel is proposed to enhance feature extraction capabilities. The red box in Fig.~\ref{fig: 11} depicts the spectral convolution kernel. The upper part of the feature map represents MFSC, while the lower part represents SF. Since MFSC and SF are temporally aligned, the convolution kernel can learn spectral representations on specific time slices. For instance, if the convolution kernel matches a specific window feature during the sliding process, the corresponding clause is satisfied, and its weight is output. This behavior matches the definition of the CTM model introduced in Section \ref{subsec: 2.1}, thus effectively achieving keyword feature recognition while ensuring interpretability. Furthermore, a larger convolution kernel inherently reduces the number of sliding operations, which translates to lower hardware overhead in the context of this work. For further trade-off analysis between convolution kernel and model size, please refer to the Appendix.

%\subsubsection{State-driven Architecture}
\subsubsection{State-Driven Architecture}

As mentioned in Section \ref{subsec: 2.2}, in the data-driven architecture, the systolic array brings efficient data reuse and high parallelism but lacks flexibility in processing sparse data. In the event-driven architecture, the event-driven strategy can fully utilize the sparsity of input activations, although its parallelism is limited by the unpredictability of spiking events. Benefiting from the sparse determinism and high sparsity of the CTM, we propose a state-driven architecture that addresses the inherent limitations of the above two architectures.

\begin{figure}[t]
    \centering
    \includegraphics[width=0.9\linewidth]{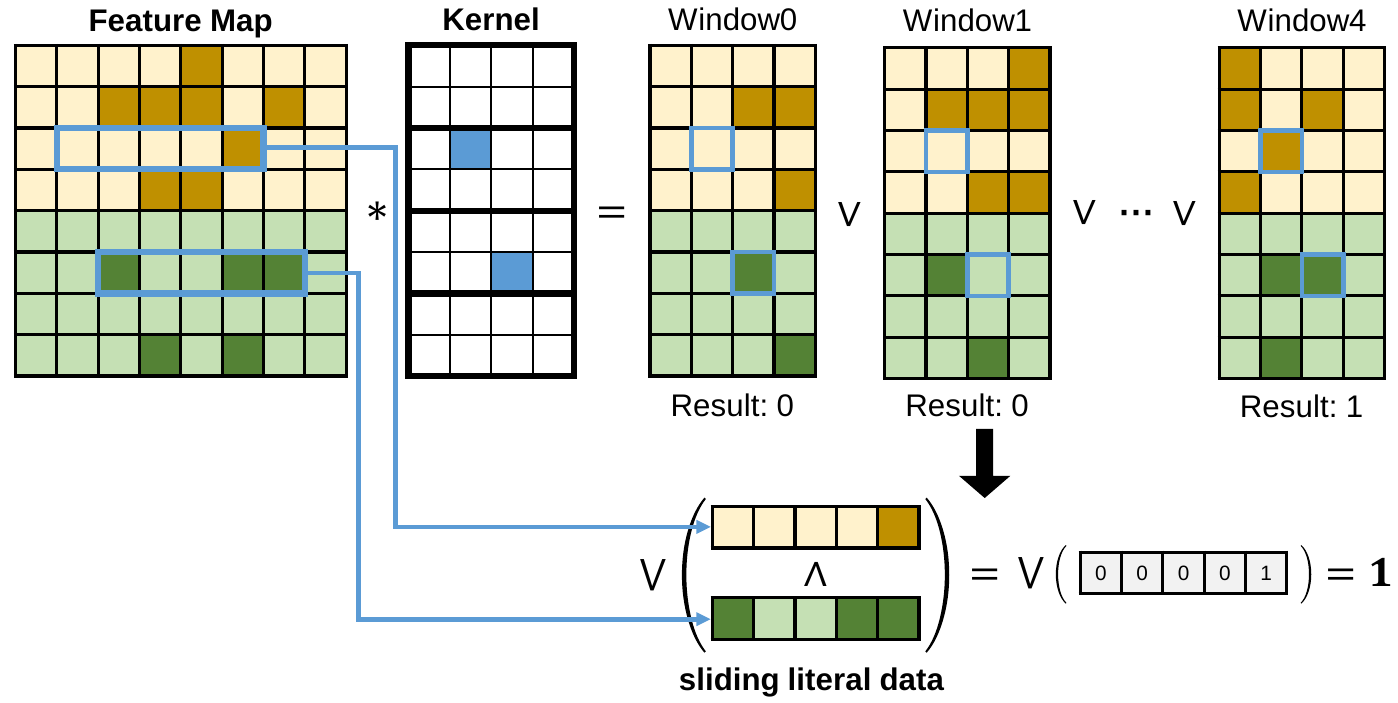}
    \caption{The computation process of the CTM and the state-driven strategy. The sliding literal data is obtained according to the position of included TAs, and it is directly computed to equivalently achieve the convolution result.}
    \label{fig: 12}
\end{figure}

Fig.~\ref{fig: 12} describes the convolution process and the state-driven strategy. The convolution kernel slides across the feature map to detect specific patterns. The result of the window is obtained by performing logical operations between the feature map and the kernel. In CTMs, the excluded TA represents ``don't care", so we only need to focus on the feature map data corresponding to the included TA positions, i.e., the blue square in the window. When all the feature map data corresponding to the positions of blue squares are 1, it indicates that the window has a specific pattern. As long as the specific pattern is identified in any one window, the clause is satisfied. We found that for each included TA, the required data is the feature data covered by it during the sliding process. Therefore, the sliding process is unrolled to obtain our state-driven strategy. 

Firstly, the row data of the feature map is read based on the row index of the included TA. Secondly, the actual covered data in the row data is selected based on the column index of the included TA. The data covered during sliding window operations is called sliding literal data. The length of sliding literal data is equal to the number of windows. Thirdly, logical AND operations are performed on the sliding literal data of each included TA to obtain the result for each sliding window. This step is performed in the PE array, so each PE only contains one AND gate. Finally, as long as the result of one sliding window is 1, the clause is satisfied. Thanks to the OG-BCSR algorithm, the index of included TAs can be obtained through decompression. When the included TAs are decompressed sequentially, the hardware architecture for a CTM with N windows is equivalent to an N$\times$1 PE array.

\begin{figure}[t]
    \centering
    \includegraphics[width=1.0\linewidth]{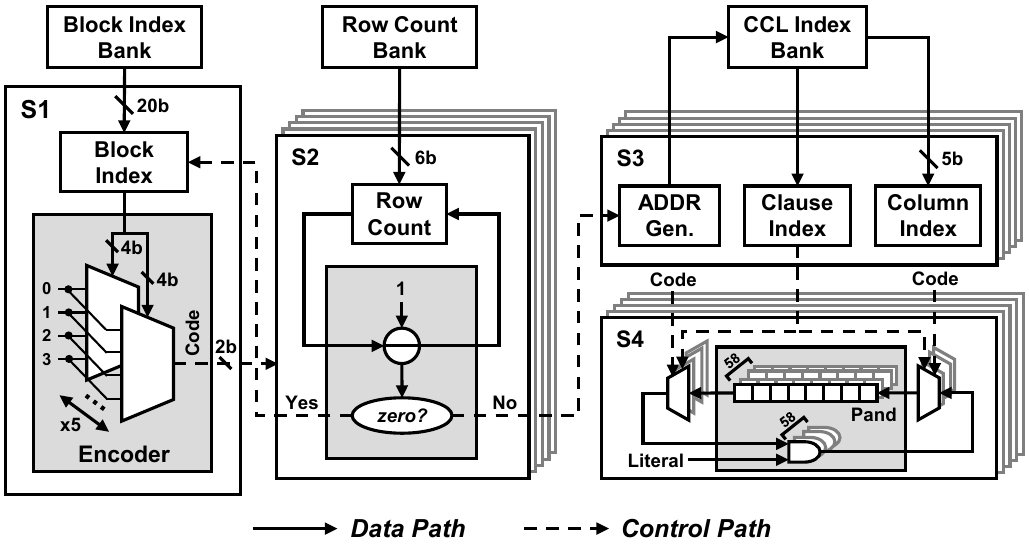}
    \caption{The core pipeline architecture in TsetlinKWS. The OG-BCSR decompression circuit comprises three stages (S1–S3), followed by the PE array (S4). The `×5' notation denotes five parallel priority encoding units.}
    \label{fig: 13}
\end{figure}

To achieve horizontal reuse of input features, we employ a priority encoding scheme for decompression, rather than simply duplicating the hardware for parallel matrix decompression. This decision is motivated by two concerns: hardware overhead and utilization. Firstly, unrolling all sliding windows in the state-driven strategy requires multiplexers, which means that simply duplicating the hardware would introduce substantial hardware overhead. Secondly, to avoid access conflicts in the input feature bank, the matrix decompression process is performed synchronously: the next block begins processing only after finishing the decompression of all included TAs in the current block. Due to the high sparsity of the TA action matrix, synchronous decompression of multiple matrices results in low PE utilization. For instance, if only 6 out of 20 matrices contain included TAs in the current block, the utilization of the decompression logic and PEs falls below 30\%.

Fig.~\ref{fig: 13} illustrates the pipeline architecture of the decompression and computation circuit. In stage S1, the block indices of 20 TA action matrices are loaded into registers. The priority encoder groups every four matrices and generates five 2-bit ``Code" signals, each identifying the active matrix within its group. In stage S2, each ``Code" signal is used to select the corresponding row count value from its assigned TA matrix. This value is then decremented each cycle to decompress included TAs sequentially. Stage S3 is activated whenever the row count value is non-zero, reading out the corresponding column and clause indices. Once the row count value reaches zero, the block index is updated to proceed to the next matrix. In stage S4, the partial logic AND (Pand) result is updated according to the ``Code" signal and clause index. By applying the priority encoding method, 20 matrices are time-shared and allocated across five parallel decompression logic units and PE columns, effectively alleviating resource idleness.

%\subsubsection{Two-stage Scheduling Optimization Algorithm}
\subsubsection{Two-Stage Scheduling Optimization}

\begin{algorithm}[t]
\caption{Two-Stage Scheduling Optimization Algorithm.}\label{alg:alg1}
\textbf{Input}: Initial solution $ \mathbf{s} \in \mathbb{R}^{p \times n \times r} $ (parallelism $p=5$, size $n=4$, and rounds $r=3$), Action count matrix $ \mathbf{E} \in \mathbb{R}^{pnr \times b} $ (blocks $b=32$), Initial temperature $T_0$, Cooling coefficient $\alpha \in (0,1)$, Termination temperature $ T_{min} $.\\
\textbf{Output}: Optimized scheduling solution $\mathbf{s}_{opt}$.\\
\vspace{-0.5cm}
\begin{algorithmic}[1]
%\STATE \textbf{Input}: Initial solution $ \mathbf{s} \in \mathbb{R}^{p \times n \times r} $ (parallelism $p=5$, size $n=4$, and rounds $r=3$), Action count matrix $ \mathbf{E} \in \mathbb{R}^{pnr \times b} $ (blocks $b=32$), Initial temperature $T_0$, Cooling coefficient $\alpha \in (0,1)$, Termination temperature $ T_{min} $.
%\STATE \textbf{Output}: Optimized scheduling solution $\mathbf{s}_{opt}$.

\item[]
\STATE // Cost function of cycle estimation.
\STATE \textbf{Function} \textsc{Cost}($\mathbf{s}, \mathbf{E}$):
\STATE \quad $cycle \gets 0$
\STATE \quad \textbf{for} $i \gets 1$ \textbf{to} $r$ \textbf{do}
\STATE \quad \quad Obtain the $i$-th round indices: $\mathbf{idx} \gets \mathbf{s}_{:,:,i}$
\STATE \quad \quad Select subset $\mathbf{V} \in \mathbb{R}^{pn \times b}$: $\mathbf{V} \gets \mathbf{E}(\text{vec}(\mathbf{idx}))$
%$\mathbf{V}\gets \mathbf{E}_{\text{vec}(\mathbf{idx}), :}$
\STATE \quad \quad \textbf{for} $j \gets 1$ \textbf{to} $b$ \textbf{do}
\STATE \quad \quad \quad Reshape $\mathbf{V}_{:,j}$ to $\mathbf{G} \in \mathbb{R}^{p \times n}$: $\mathbf{G} \gets \text{reshape}(\mathbf{V}_{:,j})$
\STATE \quad \quad \quad $cycle \gets cycle + \max_{k \in \{1,\dots,p\}} \sum_{m=1}^{n} \mathbf{G}_{k,m}$
\STATE \quad \quad \textbf{end for}
\STATE \quad \textbf{end for}
\STATE \quad \textbf{return} $cycle$
%\STATE \textbf{end Function}

\item[]
\STATE /* Simulated Annealing Optimization. */
\FOR{$stage \in \{1, 2\}$}
    \STATE // 1: Pre-optimization, 2: Post-optimization.
    \STATE \textbf{Initialize}: Current temperature $T \gets T_0$.
    \WHILE{$T > T_{min}$}
        \IF{$stage = 1$}
            \STATE Randomly swap elements between two groups to obtain a new solution $\mathbf{s}_{new}$.
        \ELSIF{$stage = 2$}
            \STATE Randomly swap elements within each group to obtain a new solution $\mathbf{s}_{new}$.
        \ENDIF
        \STATE // Calculate energy difference.
        \STATE $\Delta E \gets \textsc{Cost} (\mathbf{s}_{new}, \mathbf{E}) - \textsc{Cost}(\mathbf{s}, \mathbf{E})$
        \IF{$\Delta E < 0$ \textbf{or} $\exp(-\Delta E/T) > \text{rand}(0,1)$}
            \STATE $\mathbf{s} \gets \mathbf{s}_{new}$
        \ENDIF
        \STATE $T \gets \alpha \cdot T$
    \ENDWHILE
\ENDFOR
\STATE $\mathbf{s}_{opt} \gets \mathbf{s}$
\STATE \textbf{return} $\mathbf{s}_{opt}$

\end{algorithmic}
\end{algorithm}

The sparsity of TM models is unstructured, which presents a key hardware challenge in achieving load balancing under high parallelism. For event-driven architectures, due to the unpredictability of spiking events, additional load-balancing hardware is necessary, such as special dispatchers and rotating FIFOs. However, these load-balancing components are undesirable for lightweight edge KWS accelerators since they introduce additional memory and control complexity. Thanks to the deterministic sparsity of TMs, the activity of the hardware can be simulated in software, representing the possibility of offline load balancing.

The purpose of load balancing is to improve hardware utilization and reduce computation cycles. Specifically, we pre-calculate an action count matrix $\mathbf{E}$, which records the number of non-zero elements in each block of the TA matrices, as the basis for cycle estimation. We then utilize the simulated annealing algorithm to achieve optimization, with the goal of minimizing the total cycles. Due to the enormous search space, the optimization process is divided into two stages to progressively reduce inter-round and intra-round workload imbalance. For the 120-clause CTM model, the 60 TA matrices of each class are mapped onto $p=5$ PE columns and executed over $r=3$ rounds. In the first stage, matrices are randomly exchanged between different rounds to mitigate inter-round imbalance. In the second stage, matrices are exchanged between PEs within the same round to alleviate intra-round load variance. Algorithm \ref{alg:alg1} summarizes the proposed optimization process.

\begin{figure}[t]
    \centering
    \includegraphics[width=0.95\linewidth]{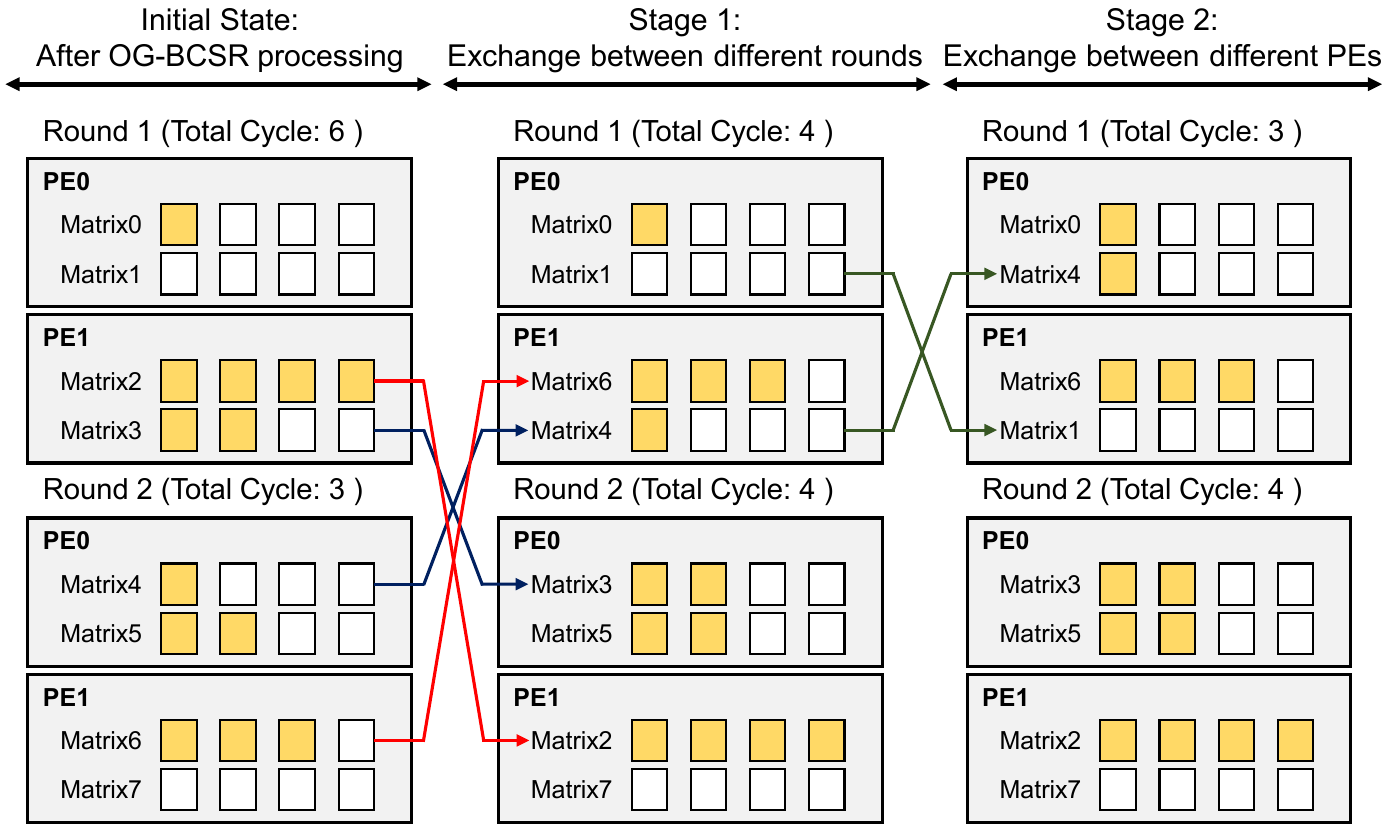}
    \caption{Example of a two-stage scheduling optimization strategy.}
    \label{fig: 14}
\end{figure}

\begin{figure}[t]
    \centering
    \includegraphics[width=0.9\linewidth]{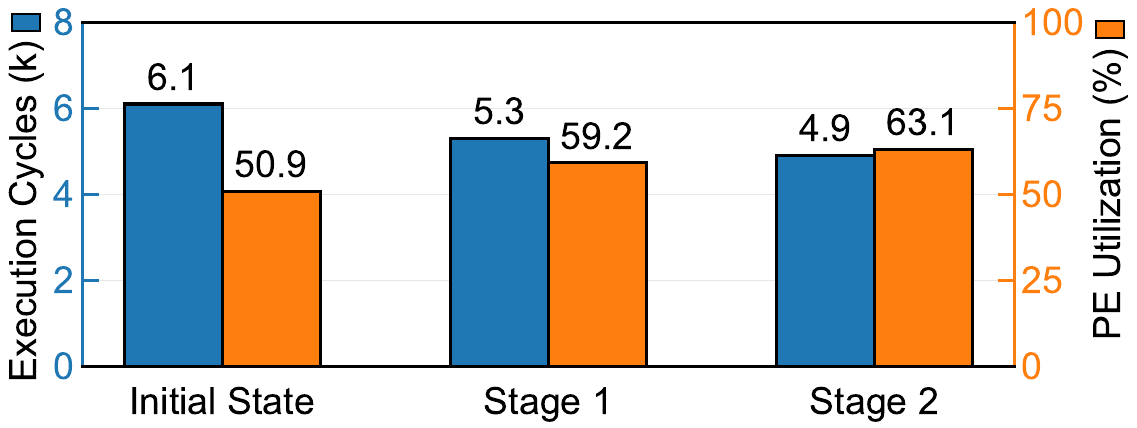}
    \caption{Comparisons of the execution cycle and PE utilization in each optimization stage.}
    \label{fig: 15}
\end{figure}

Fig.~\ref{fig: 14} depicts a simplified example of the two-stage scheduling optimization strategy. In Stage 1, the imbalance between different rounds is alleviated after swapping. In Stage 2, the matrix assignment among PEs within each round is further refined to improve intra-round PE utilization. Through this two-stage optimization, the idle cycles of PE0 in round 1 are reduced from 5 to 1, leading to a shorter overall execution latency. We evaluate the two-stage scheduling optimization algorithm on the 120-clause CTM model. Fig.~\ref{fig: 15} shows the comparison of the execution cycles and PE utilization obtained by software simulations at each optimization stage. After the two-stage optimization, PE utilization increases to 63.1\%, saving 1.2k execution cycles. The optimized schedule enables the accelerator to satisfy the target inference latency of 16 ms at a clock frequency of 400 KHz.

\subsubsection{Theoretical Performance Analysis}

To evaluate the effectiveness of the proposed state-driven architecture, we compare the theoretical memory-access costs and inference latencies of three accelerator architectures under the same workload. The analysis considers $C=12$ classes, $Cl=120$ clauses per class, input dimensions $(D,T)=(64,64)$, and convolution kernel dimensions $(M,N)=(64,7)$. Let $Sp$ denote the fraction of included TAs, such that the TA sparsity is $1-Sp$. Following \cite{granmo2019convolutional}, the number of TAs associated with each clause is approximated as
\begin{equation}
    N_{\mathrm{TA}}=M\times(N+1)\times2=1024.
\end{equation}

We quantify memory-access overhead using a normalized memory-access cost:
\begin{align}
    \mathcal{MC}_{\mathrm{mem}} &= \mathcal{MC}_i + \lambda_w \mathcal{MC}_w + \mathcal{MC}_o  \nonumber\\
                                &= a\times B_i + \lambda_w b\times B_w + c\times B_o.
    \label{eq:mem_cost}
\end{align}
where $\mathcal{MC}_i$, $\mathcal{MC}_w$, and $\mathcal{MC}_o$ represent the access costs of input activations, weights, and output activations, respectively. $a$, $b$, and $c$ are the corresponding access counts, while $B_i$, $B_w$, and $B_o$ denote the corresponding access bit-widths. Since weight storage requires a larger SRAM capacity and incurs higher access energy than input and output activations, a scaling factor of $\lambda_w=3$ is considered \cite{chen2016eyeriss}. The inference latency $L$ is calculated based on the number of processing steps and the throughput of each architecture.

\textbf{Naïve data-driven architecture.}
To maximize weight reuse, we assume a weight-stationary dataflow. The on-chip weight SRAM provides a bandwidth of 1024 bits/cycle, allowing the TA data associated with one clause to be loaded in a single cycle. Specifically, each clause accesses the input features across all $D$ channels and $T$ temporal positions. It requires $T-N+1$ computation cycles to process all sliding windows, with each output position involving one write operation. Under these assumptions, the normalized memory-access cost and inference latency are given by
\begin{align}
    \mathcal{MC}_i^{(D)} &= (C\times Cl)\times (D\times T), \nonumber\\
    \mathcal{MC}_w^{(D)} &= (C\times Cl)\times N_{\mathrm{TA}}, \nonumber\\
    \mathcal{MC}_o^{(D)} &= (C\times Cl) \times (T-N+1), \\
    L^{(D)} &= (C\times Cl)\times (T-N+1)~\text{cycles}. \nonumber
\end{align}

\textbf{Event-driven architecture.}
We estimate the performance of the event-driven architecture based on the output-stationary dataflow in SpinalFlow \cite{narayanan2020spinalflow}. For each feature row, loading the corresponding kernel row across all classes requires $C\times Cl\times (N+1)\times 2$ bits. Transferring these data in a single cycle would require an impractically high memory bandwidth. We therefore process the classes sequentially, assuming an on-chip weight-SRAM bandwidth of 1920 bits/cycle, which allows one class-specific kernel row to be loaded per cycle. We assume that input events are streamed in row-major order, allowing each kernel row to be loaded once and reused across the corresponding input events. Furthermore, an output scratchpad of $Cl\times N$ bits is employed to temporarily store and reuse intermediate output activations. Each output activation update requires one read and one write operation, resulting in $Cl\times N\times 2$ memory accesses. The number of input events is
\begin{equation}
N_{\mathrm{ev}}=D\times T\times 2\times \eta=D\times T.
\end{equation}
where the factor of 2 accounts for the positive and negative literals, and \(\eta=50\%\) represents their average activity factor. Under the event-driven paradigm, we have
\begin{align}
    \mathcal{MC}_i^{(E)} &= C\times N_{\mathrm{ev}}, \nonumber\\
    \mathcal{MC}_w^{(E)} &= (C\times M)\times (Cl\times (N+1)\times 2), \nonumber\\
    \mathcal{MC}_o^{(E)} &= (C\times N_{\mathrm{ev}})\times (Cl\times N\times 2), \\
    L^{(E)} &= C\times N_{\mathrm{ev}}~\text{cycles}. \nonumber
\end{align}

\textbf{State-driven architecture.}
The proposed architecture adopts a weight-stationary dataflow and exploits a scheduling parallelism of $p$ to enable input-activation reuse. As shown in Fig.~\ref{fig: 15}, 20 weight matrices are time-multiplexed across $p=5$ PE columns, achieving an average PE utilization of $U=63.1\%$. With $p$-way scheduling, input accesses are amortized across the parallel PEs, while each included TA requires a 6-bit column-index access and contributes to $T-N+1$ output positions. Let $N_{\mathrm{incl}}=C\times Cl\times N_{\mathrm{TA}}\times Sp$ denote the total number of included TAs. Based on these strategies, the memory-access costs and inference latency are given by
\begin{align}
    \mathcal{MC}_i^{(S)} &= \frac{ N_{\mathrm{incl}}}{p}\times T, \nonumber\\
    \mathcal{MC}_w^{(S)} &= N_{\mathrm{incl}}\times 6, \nonumber\\
    \mathcal{MC}_o^{(S)} &= N_{\mathrm{incl}}\times (T-N+1)\times 2, \\
    L^{(S)} &= \frac{ N_{\mathrm{incl}}}{p\times U}~\text{cycles}. \nonumber
\end{align}

\begin{figure}[t]
    \centering
    \includegraphics[width=\linewidth]{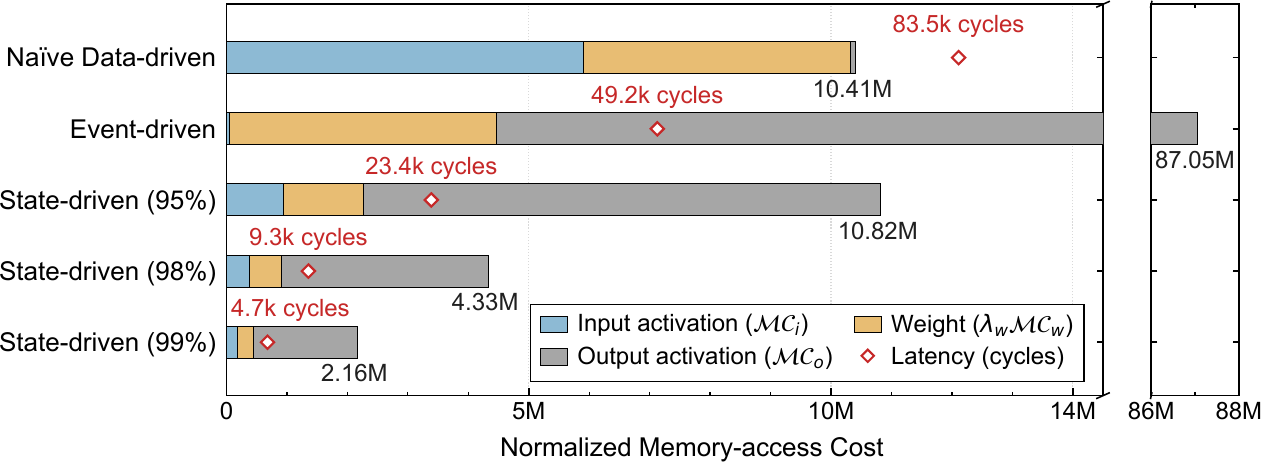}
    \caption{Comparison of normalized memory-access costs (decomposed into input activations, weights, and output activations) and inference latencies across different architectures or TA sparsity levels.}
    \label{fig: memory_latency_comparison}
\end{figure}

Fig.~\ref{fig: memory_latency_comparison} compares the theoretical normalized memory-access costs and inference latencies of the three architectures. The naïve data-driven architecture incurs substantial latency because it processes the convolutional workload without exploiting sparsity. The event-driven architecture reduces input activation accesses by processing only active input events. However, its output-stationary dataflow requires frequent intermediate output activation updates, resulting in increased output-access overhead. In contrast, the proposed state-driven architecture combines compressed weight encoding, weight-stationary dataflow, and parallel scheduling to exploit TA sparsity and improve data reuse. At typical TA sparsity levels of 98\%--99\%, the proposed architecture achieves the lowest normalized memory-access cost and inference latency. When the sparsity rate decreases to 95\%, its latency advantage persists, but the total normalized memory-access cost exceeds that of the naïve data-driven architecture. These results highlight the importance of high TA sparsity to the effectiveness of the proposed architecture.

\section{Implementation and measurement Results}\label{sec: 4}

\subsection{Experimental Setup and Performance Measurement}
\begin{figure}[t]
    \centering
    \includegraphics[width=0.9\linewidth]{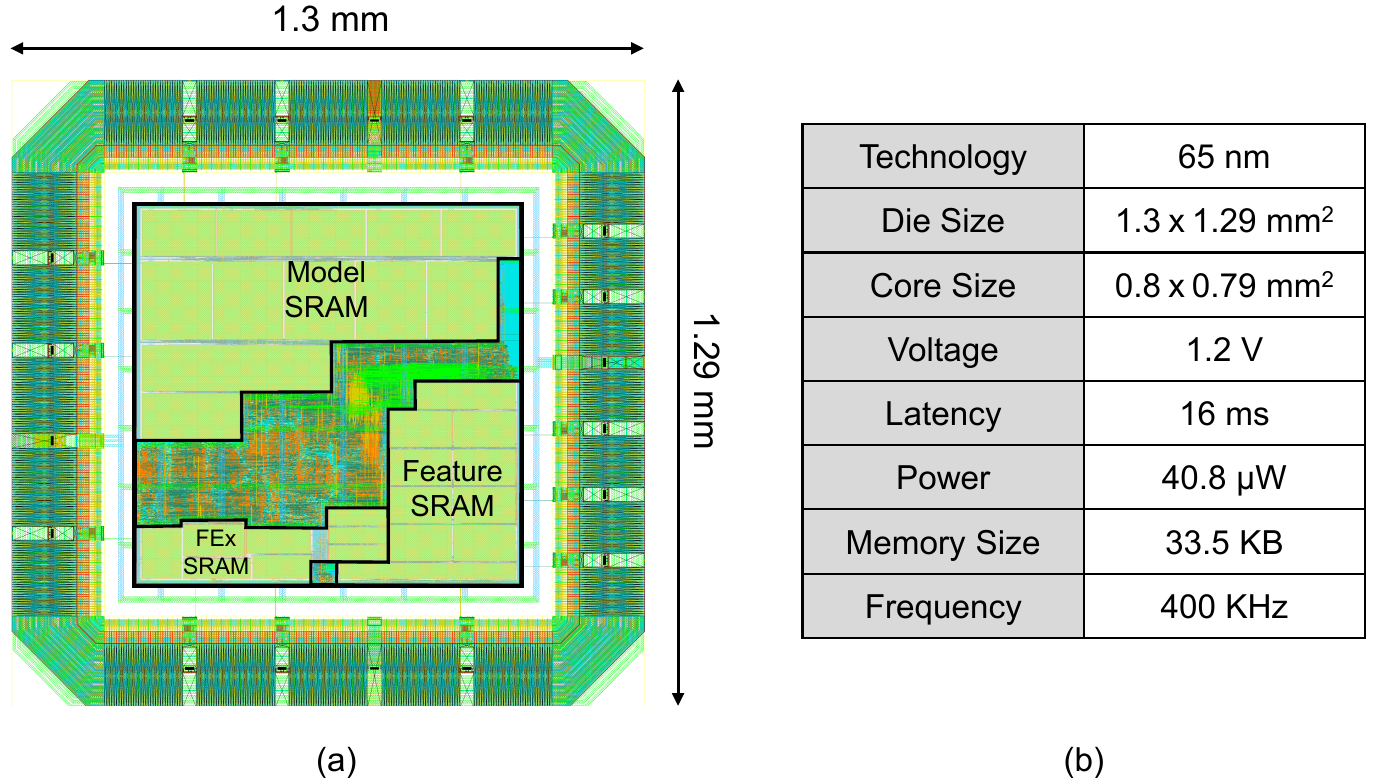}
    \caption{(a) The layout of TsetlinKWS (exported from Cadence Virtuoso). (b) The performance metrics of TsetlinKWS.}
    \label{fig: 16}
\end{figure}

\begin{figure}[t]
    \centering
    \includegraphics[width=1\linewidth]{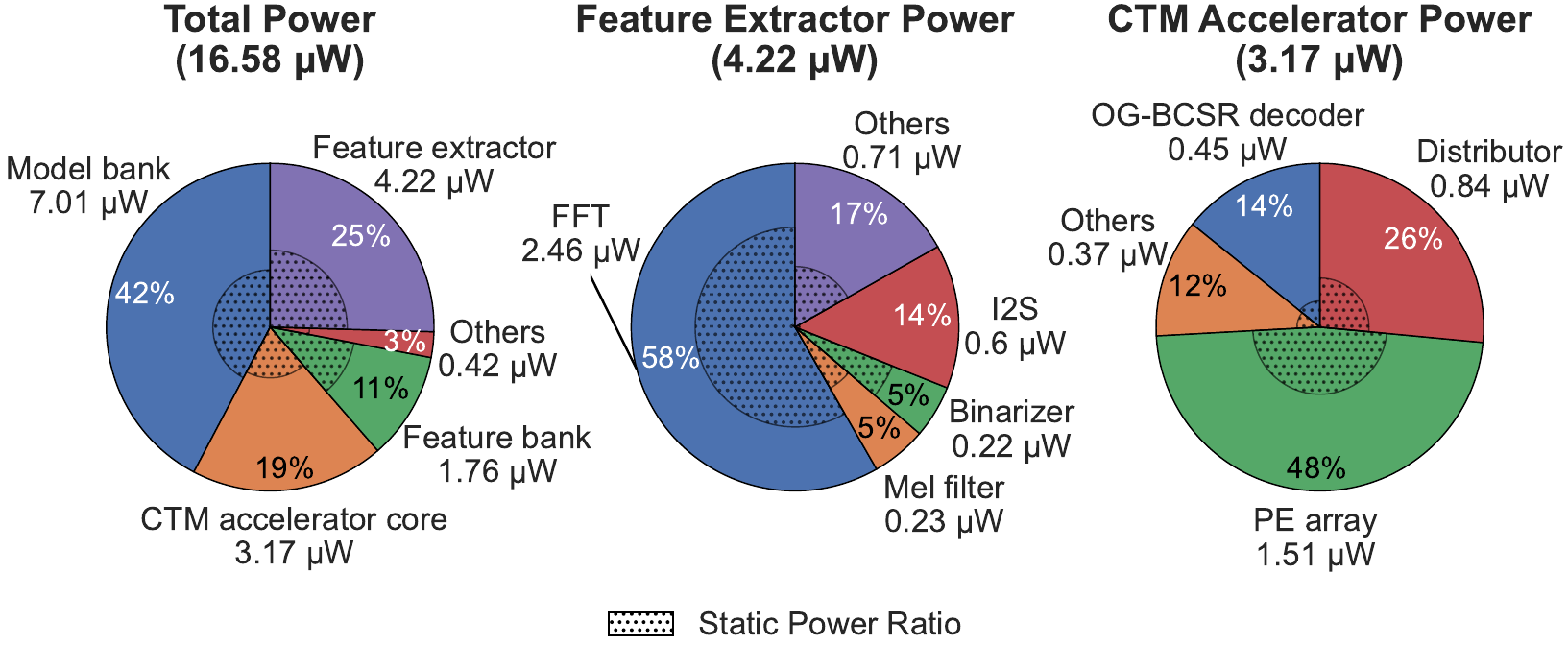}
    \caption{Power breakdown of TsetlinKWS, feature extractor, and CTM accelerator at 0.7 V (measured on Synopsys PTPX).}
    \label{fig: 17}
\end{figure}

In this work, the model parameters of TsetlinKWS are all stored on-chip by the SPI interface during initialization. The model is trained using the Google Speech Commands Dataset (GSCD) \cite{warden2018speech}, which contains 35 keywords. Among them, ten command keywords are trained as primary keywords, including ``Yes", ``No", ``Up", ``Down", ``Left", ``Right", ``On", ``Off", ``Stop", and ``Go". Additionally, there are two categories: ``Silence" (obtained from the background noise file) and ``Unknown" (comprising the remaining 25 keywords), which are also included. During training, we add white noise \cite{varga1993assessment} to the keyword audio at a randomly selected SNR between 20 and 40 dB, since the current single-layer CTM architecture is primarily evaluated under moderate noise conditions. Additionally, data augmentation is applied by randomly altering the audio pitch. The training, validation, and test sets contain 74922, 4490, and 4870 audio examples, respectively. The software code of the MFSC-SF pre-processing is designed to ensure numerical consistency with its hardware counterpart. For the 120-Clause CTM model, the hyperparameters $T$ and $s$ are set to 300 and 8.0, respectively. The detailed hyperparameter selection rationale is provided in the Appendix.

The proposed design is implemented and evaluated under TSMC 65 nm LP CMOS technology. Clock gating and operand isolation techniques are applied to further reduce power consumption. Fig.~\ref{fig: 16}(a) and Fig.~\ref{fig: 16}(b) show the layout and measurement results of TsetlinKWS, respectively. Due to the lack of a low-voltage library, measurements are conducted at 25 °C and 1.2 V, with all reported results excluding IO Pad power and scan chains. The core power at 1.2 V is 40.8 \textmu W. The die area and core area are 1.67 mm$^2$ and 0.63 mm$^2$, respectively. On-chip memory includes feature extraction SRAM (1.53 KB), model SRAM (27.25 KB), and feature SRAM (4.75 KB). The proposed system performs feature extraction and inference at a frequency of 400 KHz. Additionally, an 8.192 MHz I2S master clock is divided to generate a 1.024 MHz bit clock and a 16 KHz frame clock, which controls the external I2S digital microphone module for audio data acquisition.

\definecolor{Mercury}{rgb}{0.894,0.894,0.894}
\begin{table*}
\centering
\caption{Comparison of SoTA KWS Works}
\label{tab: 1}
\resizebox{\textwidth}{!}{
\begin{tblr}{
  colspec = {c c c c || c c},
  cells = {c, m},
  row{1} = {Mercury},
  row{2} = {Mercury},
  cell{1}{1} = {r=2}{},
  cell{1}{2} = {c=3}{},
  cell{1}{5} = {c=2}{},
  cell{3}{1} = {Mercury},
  cell{4}{1} = {Mercury},
  cell{5}{1} = {Mercury},
  cell{6}{1} = {Mercury},
  cell{7}{1} = {Mercury},
  cell{8}{1} = {Mercury},
  cell{9}{1} = {Mercury},
  cell{10}{1} = {Mercury},
  cell{11}{1} = {Mercury},
  cell{12}{1} = {Mercury},
  cell{13}{1} = {Mercury},
  cell{14}{1} = {Mercury},
  cell{15}{1} = {Mercury},
  cell{16}{1} = {Mercury},
  cell{17}{1} = {Mercury},
  cell{18}{1} = {Mercury},
  cell{19}{1} = {Mercury},
  cell{20}{1} = {Mercury},
  cell{21}{1} = {Mercury},
  vlines,
  hline{1,3-22} = {-}{},
  hline{2} = {2-6}{},
}
~ ~                                                        & \textbf{ Batch-based Accelerator } &                                                              &                                          & \textbf{ Frame-based Accelerator }                     &                                         \\
                                                           & \textbf{ This work }               & \textbf{ OJSSCS'23 \cite{liu20233} }                                     & \textbf{ TCAS-I'26 \cite{li20262} }                  & \textbf{ JSSC'20 \cite{giraldo2020vocell} }                                & \textbf{ TCASAI'24 \cite{chen2024deltakws} }               \\
\textbf{ Technology }                                      & 65 nm                               & 22 nm                                                         & 28 nm                                     & 65 nm                                                   & 65 nm                                    \\
\textbf{ Voltage (V) }                                     & 0.7                                & 0.54                                                          & 0.42                                      & 0.6                                                    & 0.6/0.65                                \\
\textbf{ Core Area (mm\textsuperscript{2}) }               & {0.63 \\ (FEx: 50\%, Accel: 50\%)}                              & {0.6 \\ (FEx: 42\%, Accel: 58\%)}                             & {0.37 \\ (FEx: 16.8\%, Accel: 83.2\%)}                                      & {2.56 \\ (Accel: 27\%, Other: 73\%)}                                                   & {0.78 \\ (FEx: 11\%, Accel: 89\%)}                                    \\
\textbf{ N. Core Area\textsuperscript{a} (mm\textsuperscript{2}) }               & \textbf{0.63}                               & 5.56                                                        & 2.15                                      & 2.56                                                   & \underline{0.78}                                    \\
\textbf{ Sample Rate }                                     & 16 KHz                              & 8 KHz                                                        & 8 KHz                                    & 16 KHz                                                  & 8 KHz                                    \\
\textbf{ System Frequency }                                & 400 KHz                             & 250 KHz                                                       & 64 KHz                                   & 250 KHz $\sim$ 8 MHz                                            & 125 KHz                                  \\
\textbf{ Power }                                           & 16.58 \textmu W                             & 3.8 \textmu W                                                        & 2.2 \textmu W                                   & 16.11 \textmu W                                                 & 7.36 \textmu W ($\Delta _{TH} = 0$)                \\
{\textbf{Inference-Core Power}\\\textbf{(Logic + Memory) }} & {
  11.94 \textmu W\\(3.17 \textmu W + 8.77 \textmu W)
  }      & { 2.68 \textmu W\\(2.11 \textmu W + 0.57 \textmu W) }                                                        & {
  1.69 \textmu W\\(0.31 \textmu W + 1.38 \textmu W)
  }              & {6.01 \textmu W \\(4.54 \textmu W + 1.47 \textmu W) }                                                   & {
  6.14 \textmu W \\(4.67 \textmu W + 1.47 \textmu W)
  }            \\
{\textbf{N. Inf.-Core Power}\\\textbf{(Logic + Memory)\textsuperscript{a}}} & {
  11.94 \textmu W\\(\underline{3.17} \textmu W + 8.77 \textmu W)
  }      & { 5.58 \textmu W\\(4.40 \textmu W + 1.18 \textmu W) }                                            & {
  3.02 \textmu W\\(\textbf{0.55} \textmu W + 2.47 \textmu W)
  }              & {6.01 \textmu W \\(4.54 \textmu W + 1.47 \textmu W) }                                                  & {
  6.14 \textmu W \\(4.67 \textmu W + 1.47 \textmu W)
  }            \\
\textbf{ Latency (ms) }                                    & \textbf{16}                                 & \textbf{16}                                                           & \textbf{16}                                      & \textbf{16}                                                     & \underline{16.4} ($\Delta _{TH} = 0$)                  \\
\textbf{ N. Inf.-Core Energy\textsuperscript{a} }               & 191 nJ                              & 89.3 nJ                                                       & 48.3 nJ                                     & 96.2 nJ                                                  & 100.7 nJ ($\Delta _{TH} = 0$)                               \\
\textbf{ N. PPA Product\textsuperscript{a} }                                           & \textbf{60.2}          & 287.9    & 86.4  & \underline{66.5}     & 69.9                             \\
\textbf{ Model }                                           & CTM                                & TWN                                                          & BC-ResNet                                & LSTM                                                   & $\Delta$GRU                             \\
\textbf{ Model Size }                                      & 19.9 KB                             & 5.3 KB                                                        & 7.6 KB                                   & 10.1 KB\textsuperscript{b}                              & 14.8 KB                                  \\
\textbf{ On-chip Memory }                                  & 33.5 KB                             & 14 KB                                                       & 9 KB                                   & 96 KB                                                   & 25.7 KB                                  \\
\textbf{ Dataset }                                         & GSCD
  (12)                        & GSCD
  (12)                                                  & GSCD
  (12)                              & GSCD
  (12)                                            & GSCD
  (12)                             \\
\textbf{ \#Operations/Inference }                               & 907k
  AND                         & 4.93m
  16b Approx. ADD                                         & {
  1.17m MUL + 1.26m ADD
  } & {
  20k
  8*8b MUL \\+ 20k 32b ADD\textsuperscript{b}} & {
  15k
  8*12b MUL \\+ 15k 12b ADD
  } \\
\textbf{ \#Norm\_Ops./Inf.\textsuperscript{c}}       & \textbf{907k}                               & 374.7m                                                         & 250.7m                                     & 9.8m\textsuperscript{b}                                & \underline{9.3m} ($\Delta _{TH} = 0$)                                    \\
\textbf{ Accuracy }                                        & {87.35\% @clean \\ 83.10\% @20dB}                      & {
  90.6\%
  @clean
  \\85.4\%
  @5dB
  } & {
  94.3\%
  @clean
  \\87.6\%
  @0dB
  }   & 90.87\% @clean                                          & 89.5\% @clean                            
\end{tblr}

}

\medskip
\begin{minipage}{\linewidth}
\footnotesize
\textsuperscript{a} ``N.'' denotes that the core area, inference-core power, energy, and PPA product are normalized to 65 nm using the \textit{DeepScaleTool} \cite{sarangi2021deepscaletool}.\\
\textsuperscript{b} Assuming the structure of the LSTM model is: input(13)-hidden(64)-output(12). The model parameters are 4 bits, and are restored to 8 bits during calculation.\\ 
\textsuperscript{c} Normalize the number of KWS operations per inference to 2-input AND gate operations.
\end{minipage}

\end{table*}

For KWS applications, the operating voltage can generally be scaled down below 0.7 V without timing violations, as the low operating frequency provides sufficient timing margin. To facilitate fair comparison with other works, we scale dynamic power by the square of the voltage ratio and static power by the voltage ratio. This is a conservative estimate for leakage power, as it does not account for the additional reduction in leakage current at lower voltages. The core voltage is scaled from 1.2 V to 0.7 V for this analysis.

Fig.~\ref{fig: 17} shows the power breakdown at 0.7 V. After voltage scaling, the core power is decreased from 40.8 \textmu W to 16.58 \textmu W. The 27.25 KB SRAM used for storing the CTM model accounts for 42\% of the total power. The power consumptions of the feature extractor and feature bank are 4.22 \textmu W and 1.76 \textmu W, respectively. Within the feature extractor, the FFT module dominates the power consumption. This is because other modules employ various low-power techniques to reduce computational complexity, whereas the FFT scheme adopted here is not the optimal choice for real-valued audio signal processing. The optimization of the FFT is not included in this work. For the CTM accelerator, the majority of the power is consumed by the PE array, including its internal 2320 Pand registers, as expected. This overhead is a direct result of our design goal to achieve sparse acceleration with minimal control overhead. Both the FFT and PE array exhibit relatively high leakage power ratios compared to other modules. As the leakage power estimation is conservative, actual power consumption is expected to be even lower.

\subsection{Comparison With SoTA Works}

Table~\ref{tab: 1} compares our proposed design with state-of-the-art KWS accelerators. All KWS chips are categorized into two types: batch-based accelerators and frame-based accelerators.

Among batch-based approaches, Liu et al. \cite{liu20233} proposed a Ternary Weight Network (TWN) accelerator to eliminate multiplication operations, yielding a normalized inference-core energy of 89.3 nJ at 65 nm. Although approximate addition is employed, the 374.7 million equivalent AND operations per inference impose a substantial computational burden and area overhead on the processing units, resulting in an inference-core PPA product of 287.9 under the definition in \eqref{eq:ppa}. Bernardo et al. \cite{bernardo2020ultratrail} deployed a shallow TC-ResNet, which entails 1.27 million Multiply-Accumulate (MAC) operations per inference, resulting in a 100 ms latency and an inference-core PPA product of 3141. Li et al. \cite{li20262} developed a learnable Mel filter bank algorithm along with an adaptive denoising strategy, enabling the deployed BC-ResNet to achieve 94.3\% accuracy. However, owing to the custom low-leakage memory occupying 74.6\% of the chip area, it incurs an inference-core PPA product of 86.4.

On the other hand, frame-based accelerators attain higher efficiency by processing features sequentially. Giraldo et al. \cite{giraldo2020vocell} employed a single-layer long short-term memory (LSTM) network that requires only 20k MAC operations per inference, yielding an inference-core energy of 96.2 nJ. Owing to this small computational workload, it delivers an inference-core PPA product of 66.5. Chen et al. \cite{chen2024deltakws} designed a custom memory architecture to reduce SRAM power by 6.6$\times$, achieving a competitive inference-core energy of 101 nJ. However, this improvement comes at the cost of a substantial area overhead, with the inference core occupying 89\% of the on-chip area and resulting in an inference-core PPA product of 69.9. These results show that the PPA product provides a more balanced evaluation by jointly considering energy-related efficiency and area overhead.

Compared to other batch-based accelerators, our design achieves a 1.4$\times$ reduction in the inference-core PPA product while occupying a significantly smaller core area. Building upon this efficiency, TsetlinKWS even outperforms the SoTA frame-based design, delivering a 1.1$\times$ reduction in the inference-core PPA product. Furthermore, TsetlinKWS requires the fewest gate-level operations per inference, even fewer than frame-based accelerators, which process only a single frame per inference.

\subsection{Discussion}
According to experimental results, the Convolutional Tsetlin Machine, as an emerging AI algorithm, demonstrates remarkable potential for keyword spotting tasks. In this paper, we construct a complete KWS hardware system and compare it with the SoTA neural network accelerators. On the basis of rigorously demonstrating its PPA advantage, we provide new insights for the hardware architecture design of CTMs. The following is a further explanation and analysis of the implementation outcomes:
\subsubsection{State-driven architecture}
The fundamental principle of a state-driven architecture lies in effectively fetching features based on deterministic weights. Since the convolution operation inherently relies on the spatial broadcasting of weights to features, the proposed architecture maximizes weight reuse and minimizes feature accesses for CTMs. Thus, it can also be extended to other applications such as image recognition for highly energy-efficient inference.
\subsubsection{Flexible category scaling}
The CTM model used in this work does not share parameters across categories. Therefore, when the required keyword categories are reduced, the model size will be linearly reduced, which can alleviate the current large memory overhead and computational latency. For example, when only two keywords are needed, the number of logical operations per frame will be reduced to 150k.
\subsubsection{Memory access optimization potential}
Due to the limited feature extraction capability of the single-layer CTM model, the CTM model is larger in size than NN-based models. For the power consumption of AI accelerators, besides computational logic, memory access is equally important. Therefore, techniques such as memory bank division, setting up near-end caches, and customizing dedicated memory blocks can effectively further reduce system power consumption.
\subsubsection{Robust feature extraction algorithm}
In our experiments, we found that due to the sensitivity of logic-based models to background noise, the KWS system proposed in this work is currently best suited to near-microphone scenarios. Previous work has shown that learnable filters and denoising algorithms can improve accuracy by at least 2 percentage points under both clean and high-noise conditions \cite{li20262}. In addition, bio-inspired acoustic features have been shown to offer greater robustness than MFCCs for KWS \cite{wang20219}. Combining a more robust pre-processing algorithm with the CTM could help improve both accuracy and noise robustness. We leave this direction for future work.
\subsubsection{Advantages and challenges}
Our results show that CTMs can yield competitive PPA metrics on audio tasks when compared with frame-based accelerators. Furthermore, the inherently simple learning mechanism of CTMs indicates that they are a highly promising solution for edge speech applications, which demand strict privacy and high adaptability. An important issue at present is that if the accuracy requirement exceeds 90\%, a larger CTM model size is required, which means that model size and memory access deserve more attention in future CTM research.

\section{Conclusion}\label{sec: 5}
This paper proposes a Convolutional Tsetlin Machine-based KWS chip through algorithm-hardware co-design, featuring an MFSC-SF feature extractor, the spectral convolution operation, and a state-driven architecture with fully sparse utilization and efficient data reuse. By compressing and scheduling the deterministically sparse model parameters, we minimize both the model size and the number of execution cycles. The proposed chip consumes only 16.58 \textmu W with a core area of 0.63 mm$^2$ at 65 nm. Compared to state-of-the-art batch-based and frame-based KWS accelerators, TsetlinKWS demonstrates 1.4$\times$ and 1.1$\times$ reductions in the inference-core PPA product, respectively. Furthermore, our design enables deterministic sequential access to all model memories, providing further opportunities for memory optimization. The introduction of spectral flux features effectively improves accuracy with low hardware overhead, enabling TsetlinKWS to achieve a classification accuracy of 87.35\% on the 12-keyword spotting task.

\section*{Acknowledgments}

The authors would like to acknowledge the Novel IC Exploration Facility at the Hong Kong University of Science and Technology (Guangzhou) for providing experimental support, and Ms. Zheng Yujin of Newcastle University for help with backend script writing. This work was supported by the Engineering and Physical Sciences Research Council (EPSRC), UK, under the projects \emph{MultiTasking and Continual Learning for Audio Sensing Tasks on Resource-Constrained Platforms} (EP/X01200X/1).

{\appendix[Hyperparameter selections for Convolutional Tsetlin Machines]\label{appendix}

This appendix presents supplementary experimental data to provide a detailed analysis of the hyperparameter search for the Convolutional Tsetlin Machine. For the training of CTMs, the convolution kernel size, together with the hyperparameters $s$ and $T$, plays a pivotal role in optimizing the model performance.

Although larger kernels provide a broader receptive field for global feature extraction, they inherently increase the model size. To ensure a fair evaluation, we maintain comparable model sizes across three kernel configurations by adjusting the number of clauses accordingly. As shown in Table~\ref{tab: 2}, the $64\times 7$ kernel achieves the optimal trade-off between recognition accuracy and model footprint.

\begin{table}[h]
\centering
\setlength{\extrarowheight}{2pt}
\caption{Experimental Results for Different Kernel Sizes}
\label{tab: 2}
\begin{tabular}{|c|c|c|c|}
\hline
Kernel Size                    & \#Clause/Class & \begin{tabular}[c]{@{}c@{}}Training Accuracy\\@20-40dB\end{tabular} & \begin{tabular}[c]{@{}c@{}}Test Accuracy\\@clean\end{tabular}  \\ 
\hline
$64\times 3$  & 240                  & 85.73\%                        & 86.55\%                     \\ 
\hline
$64\times 7$  & 120                  & 85.33\%                        & \textbf{87.35\%}                     \\ 
\hline
$64\times 15$ & 60                   & 82.83\%                        & 84.89\%                     \\
\hline
\end{tabular}
\end{table}

The specificity $s$ and threshold $T$ are the two most critical hyperparameters in the Tsetlin Machine. We perform a grid search over a range of values to determine the optimal hyperparameter combination for the 120-Clause CTM model with a kernel size of $64\times7$, as reported in Table~\ref{tab: 3}. Based on these results, $s=8.0$ and $T=300$ are selected as the optimal training hyperparameters for the model.

% \usepackage{tabularray}
% \UseTblrLibrary{diagbox}
\begin{table}[ht]
\centering
\caption{Grid Search Results for Hyperparameters}
\label{tab: 3}
\begin{tblr}{
  column{odd} = {c},
  column{4} = {c},
  cell{2}{1} = {r=2}{},
  cell{4}{1} = {r=2}{},
  cell{6}{1} = {r=2}{},
  hlines,
  vlines,
}
\diagbox[width=3em, height=2em, innerleftsep=4pt, innerrightsep=3pt]{$s$}{\raisebox{-1pt}{$T$}}  & ~                        & 100     & 300     & 500     \\
3.0            & Training Accuracy @20-40dB & 84.57\% & 85.00\% & 85.24\% \\
               & Test Accuracy @clean     & 85.26\% & 85.79\% & 85.48\% \\
8.0            & Training Accuracy @20-40dB & 85.34\% & 85.33\% & 85.56\% \\
               & Test Accuracy @clean     & 87.02\% & \textbf{87.35\%} & 87.04\% \\
10.0           & Training Accuracy @20-40dB & 84.83\% & 84.51\% & 84.96\% \\
               & Test Accuracy @clean     & 86.37\% & 86.37\% & 86.26\% 
\end{tblr}
\end{table}

%% \begin{table}[h]
%% \centering
%% \caption{Grid Search Results for Hyperparameters (Test Accuracy \%)}
%% \begin{tblr}{
%%   row{even} = {c},
%%   row{3} = {c},
%%   row{5} = {c},
%%   cell{1}{2} = {c},
%%   cell{1}{3} = {c},
%%   cell{1}{4} = {c},
%%   vline{-} = {2-5}{},
%%   hline{2-6} = {-}{},
%% }
%%                &     &     &     \\
%% \diagbox{s}{T} & 100 & 300 & 500 \\
%% 5.0            & 0\% & 0\% & 0\% \\
%% 8.0            & 0\% & 0\% & 0\% \\
%% 10.0           & 0\% & 0\% & 0\% 
%% \end{tblr}
%% \end{table}
}

%\begin{thebibliography}{1}
\bibliographystyle{IEEEtran}

\bibliography{TsetlinKWS_Reference}

\vfill

\end{document}